\documentclass[12pt]{iopart}

\usepackage{subcaption}
\usepackage{graphicx}
\usepackage{subfig}
\usepackage{epsfig}
\usepackage{iopams}
\newcommand{\p}{\partial}

\newcommand{\wh}[1]{\widehat{#1}}

\newcommand{\vek}[1]{\mathbf{#1}}

\newcommand{\spar}{\shortparallel} 
\newcommand{\nablan}{\nabla_{\perp}}
\newcommand{\nablap}{\nabla_{\spar}}

\begin{document}
\title[]{Isotope effect on filament dynamics in fusion edge plasmas}
\author{O~H~H~Meyer$^{1,2}$ and A~Kendl$^{1}$}
\address{$^1$Institut f\"ur Ionenphysik und Angewandte Physik, Universit\"at
  Innsbruck, 6020 Innsbruck, Austria} 
\address{$^2$Department of Physics and Technology, UiT - The Arctic University of Norway,
9037 Troms\o, Norway}
\ead{ole.meyer@uibk.ac.at}
\begin{abstract}
The influence of the ion mass on filament propagation in the scrape-off layer
of toroidal magnetised plasmas is analysed for various fusion relevant majority
species, like hydrogen isotopes and helium, on the basis of a computational
isothermal gyrofluid model for the plasma edge. 
Heavy hydrogen isotope plasmas show slower outward filament propagation and
thus improved confinement properties compared to light isotope plasmas,
regardless of collisionality regimes. Similarly, filaments in fully ionised
helium move more slowly than in deutrium. Different mass effects on the
filament inertia through polarisation, finite Larmor radius, and parallel
dynamics are identified.
\end{abstract}
\noindent{Keywords:\ }{isotope effect, plasma filament, blob, particle transport}
	
\submitto{\PPCF}
\maketitle


\section{Introduction}

In various tokamak experiments the confinement properties have been shown to
scale favourably with increasing mass of the main (fusion relevant) ion plasma
species, specifically hydrogen isotopes and helium \cite{Bessenrodt,Hawryluk,Liu,Xu}.  
The radial cross-field transport of coherent filamentary structures (commonly
denoted ``blobs'') in the scrape-off layer (SOL) of tokamaks accounts for a
significant part of particle and heat losses \cite{Boedo,LaBombard} to
the plasma facing components. Experimentally the ion mass effect on SOL filament
dynamics has been studied in a simple magnetised torus \cite{Theiler}.
Filamentary transport in tokamaks in general is an active subject of studies in
experiments, analytical theory, and by computations in two and three dimensions.

The basic properties of filamentary transport are reviewed in \cite{DIp_review}. 
Blob propagation results from magnetic drifts that polarise density
perturbations, thus yielding a dipolar electric potential $\phi$ whose 
resulting ${\bf B} \times \nabla \phi$ drift in the magnetic field ${\bf B}$ 
drives the filaments down the magnetic field gradient and towards the wall. 
The basic physics is illustrated by accounting for the current  
paths involved upon charging of the blob by the diamagnetic current: 
the closure is via perpendicular polarisation currents in the drift plane and
through parallel divergence of the parallel current \cite{Krash1}. 

Two-dimensional (2-d) closure schemes are discussed in Ref.~\cite{Krash}. 
Depending on parallel resistivity, the dominant closure path features distinct
dynamics: if closure is mainly through the polarisation current, the 2-d
cross-field properties are dominant, leading to a mushroom-cape shaped radial
propagation. For reduced resistivity the closure parallel to the magnetic
field direction in 3-d is dominant, and Boltzmann spinning leads to a more
coherently propagating structure at significantly reduced radial velocity. 

The isotope mass may have influence on the ${\bf E} \times {\bf B}$ shearing rate in the
edge region \cite{Hahm}, and flow shear in the edge region has been suggested
to be a main agent which controls blob formation \cite{Manz}. 
In addition, a finite ion temperature introduces poloidally asymmetric
propagation of blobs \cite{Madsen,Matthias}. The underlying finite Larmor
radius (FLR) effects have been found to contribute to favourable 
isotopic transport scaling of tokamak edge turbulence \cite{paper1}. 

In this work we study the isotopic mass effect on blob filament propagation by
employing an isothermal gyrofluid model so that relevant FLR contributions to
the blob evolution are effectively included, in addition to the mass
dependencies in polarisation and in parallel ion velocities.

\section{Gyrofluid model and computation} 
\label{model_section}

The present simulations on the isotopic dependence of 3-d filament and 2-d blob
propagation in the edge and SOL of tokamaks are based on the gyrofluid
electromagnetic model introduced by Scott \cite{scott05b}.
In the local delta-$f$ isothermal limit the model consists of
evolution equations for the gyrocenter densities $n_s$ and parallel velocities
$u_{s  \spar}$ of electrons and ions, where the index $s$ denotes the species with
$s \in (\mathrm{e}, i)$:

\begin{eqnarray} 
& \frac{\rmd_{s} n_s}{\rmd t} = & - \nablap u_{s \spar} + \mathcal{K} \left(
\phi_{s} + \tau_{s} n_{s} \right),  
\label{density} \\
\wh{\beta} \; \frac{\p A_{\spar}}{\p t} + \epsilon_{s} & \frac{\rmd_s u_{s \spar}}{\rmd
  t} = & - \nablap \left( \phi_{s} + \tau_{s} n_{s} \right) + 2 \epsilon_{s}
\tau_{s} \mathcal{K} \left( u_{s \spar} \right) - C J_{\spar}, 
\label{parvel} 
\end{eqnarray}
The gyrofluid moments are coupled by the polarisation equation
\begin{equation} 
\label{poleq}
\sum_{s} a_{s} \left[ \Gamma_{1} n_s + \frac{\Gamma_{0} - 1}{\tau_{s}} \phi \right] = 0,
\end{equation}
and Ampere's law
\begin{equation}
- \nablan^{2} A_{\spar} = J_{\spar} = \sum_{s} a_s u_{s \spar}.
\end{equation}
The gyroscreened electrostatic potential acting on the ions is given by
\begin{equation*}
\phi_{s} = \Gamma_{1} \left( \rho_{s}^{2} k_{\perp}^ {2}  \right) \wh{\phi}_{\vek{k}},
\end{equation*}
where $\wh{\phi}_{\vek{k}}$ are the Fourier coefficients of the electrostatic potential.
The gyroaverage operators $\Gamma_{0} (b) $ and $\Gamma_{1} (b) = \Gamma_{0}^{1/2} (b)$ correspond to
multiplication of Fourier coefficients by $I_{0}(b) e^{-b}$ and 
$I_{0}(b/2) e^{- b/2}$, respectively, where $I_{0}$ is the modified Bessel
function of zero'th order and $b = \rho_{s}^{2} k_{\perp}^ {2}$. 
We here use approximate Pad\'{e} forms with $\Gamma_{0} (b) \approx (1 +
b)^{-1}$ and $\Gamma_{1} (b) \approx (1 + b/2)^{-1}$ \cite{dorland93}. 

The perpendicular $\vek{E} \times \vek{B}$ advective and the parallel derivative operators
for species $s$ are given by 
\begin{equation*}
\frac{\rmd_{s} }{\rmd t} = \frac{\p }{\p t} + \delta^{-1} \left\{ \phi_{s}, ~ \right\},
\end{equation*}
\begin{equation*}
\nablap  = \frac{\p }{\p z} - \delta^{-1} \wh{\beta} \left\{ A_{\spar}, ~ \right\},
\end{equation*}
where we have introduced the Poisson bracket as
\begin{equation*}
\left\{ f, g \right\} = \left( \frac{\p f}{\p x} \frac{\p g}{\p y} - \frac{\p f}{\p y} \frac{\p g}{\p x}  \right).
\end{equation*}

In local three-dimensional flux tube co-ordinates $\{x,y,z\}$, $x$ is a (radial) flux-surface
label, $y$ is a (perpendicular) field line label and $z$ is the position along
the magnetic field line.
In circular toroidal geometry with major radius $R$, the curvature operator is given by
\begin{equation*}
\mathcal{K} = \omega_{B} \left( \sin z \; \frac{\p}{\p x} + \cos z \; \frac{\p}{\p y} \right),
\end{equation*}
where $\omega_{B} = 2 L_{\perp} / R$,
and the perpendicular Laplacian is given by
\begin{equation*}
\nablan^{2} = \left( \frac{\p^{2}}{\p x^{2}} + \frac{\p^{2}}{\p y^{2}} \right).
\end{equation*}
Flux surface shaping effects \cite{kendl06,kendl03} in more general tokamak or
stellarator geometry on SOL filaments \cite{riva17} are here neglected for simplicity.

Spatial scales are normalised by the drift scale $\rho_0 = \sqrt{T_\rme
  m_{i0}}/\rme B$, where $T_\rme$ is a reference electron temperature, $B$ is
the reference magnetic field strength and $m_{i0}$ is a reference ion mass,
for which we use the mass of deuterium $m_{i0} = m_\mathrm{D}$. 
The temporal scale is set to by $c_0 / L_{\perp}$, where $c_0 = \sqrt{T_\rme/m_{i0}}$,
and $L_{\perp}$ is a perpendicular normalisation length (e.g. a generalized
profile gradient scale length), so that $\delta = \rho_0 / L_\perp$ is the drift scale. 
The temporal scale may be expressed alternatively $L_\perp / c_0 = L_\perp /
(\rho_0 \Omega_{0}) = (\delta \Omega_{0})^{-1}$, with the ion-cyclotron
frequency $\Omega_{0} = \rho_0 / c_0$.  In the following we employ $\delta = 0.01$.

The main species dependent parameters are
\begin{eqnarray*}
a_{s} = \frac{Z_s n_{s0}}{n_{\rme 0}} , \quad \tau_{s} = \frac{T_{s}}{Z_s T_{\rme}}, \quad
\mu_{s} = \frac{m_{s}}{Z_s m_{i0}}, \\
\rho_{s}^{2} = \mu_{s}\tau_{s} \rho_{0}^{2}, \quad \epsilon_{s} = \mu_{s} \left( \frac{q R}{L_{\perp}} \right)^{2},
\end{eqnarray*}
setting the relative concentrations, temperatures, mass ratios and FLR scales
of the respective species. $Z_s$ is the charge state of the species $s$ with
mass $m_s$ and temperature $T_s$. 

The plasma beta parameter
\begin{equation*}
\wh{\beta} = \frac{4 \pi p_{\rme}}{B^{2}} \left( \frac{q R}{L_{\perp}} \right)^{2},
\end{equation*}
controls the shear-Alfv\'{e}n activity, and
\begin{equation*}
C = 0.51 \frac{\nu_{\rme} L_{\perp}}{c_{0}} \frac{m_\rme}{m_{i 0}} \left( \frac{q R}{L_{\perp}} \right)^{2}, 
\end{equation*}
mediates the collisional parallel electron response for $Z=1$ charged hydrogen
isotopes. The collisional response for other isotopes or ion species is discussed
further below.

\subsection*{Parallel boundary conditions}

We distinguish between two settings for parallel boundary conditions in 
3-d simulations. In the case of edge simulations a toroidal
closed-flux-surface (CFS) geometry is considered, and quasi-periodic globally
consistent flux-tube boundary conditions in the parallel direction
\cite{scott98} are applied on both state-variables $n_\rme, \phi$ and flux
variables $v_{\rme \spar}, u_{s \spar}$.   

For SOL simulations, the state variables assume zero-gradient Neumann (sheath)
boundary conditions at the limiter location, and the flux variables are given as 
\begin{eqnarray}
u_{s \spar}|_{\pm \pi} &= p_\rme|_{\pm \pi} = \pm \Gamma_d n_\rme|_{\pm \pi},\\
v_{\rme \spar} &= u_{s \spar}|_{\pm \pi} - J_\spar |_{\pm \pi}  = \pm \Gamma_d
[(\Lambda + 1) n_\rme|_{\pm \pi} - \phi|_{\pm \pi}], 
\end{eqnarray}
at the parallel boundaries $z = \pm \pi$ respectively \cite{Ribeiro05}. 
Note that in order to retain the Debye sheath mode in this isothermal model, the Debye current  
$J_\spar |_{\pm \pi} = \pm \Gamma_d (\phi - \Lambda T_\rme)$ is expressed as
$J_\spar |_{\pm \pi} = \pm \Gamma_d (\phi - \Lambda n_\rme)$ and the electron
pressure $p_\rme = n_\rme T_\rme$ is replaced by $p_\rme = n_\rme$ \cite{Ribeiro05}. 
This edge/SOL set-up and its effects on drift wave turbulence has been
presented in detail by Ribeiro \etal in Refs.~\cite{Ribeiro05,Ribeiro08}.

The sheath coupling constant is $\Gamma_d = \sqrt{(1 + \tau_i) / (\mu_i \wh{\epsilon})}$. 
The floating potential is given by $\Lambda = \Lambda_0 + \Lambda_i$, where
$\Lambda_0 = \log \sqrt{m_{i0} / (2 \pi m_\rme)}$ and $\Lambda_i = \log
\sqrt{\mu_i / (1 + \tau_i)}$. 
Here terms with the index $i$ apply only to the ion species. 
The expressions presented here are obtained by considering the finite ion
temperature acoustic sound speed, $c_i = \sqrt{(Z_i T_i + T_\rme) / m_i}$, 
instead of $c_0$ in Ref.~\cite{Ribeiro05}. This results in the additional
$\Lambda_i$, and the normalisation scheme yields the extra 
$\sqrt{(1 + \tau_i) / \mu_i}$ in $\Gamma_d$.

\subsection*{Numerical implementation}

Our code TOEFL \cite{kendl14} is based on the delta-$f$ isothermal electromagnetic
gyrofluid model \cite{scott05b} and uses globally consistent flux-tube
geometry \cite{scott98} with a shifted metric treatment of the coordinates
\cite{scott01} to avoid artefacts by grid deformation. In the SOL region a
sheath boundary condition model is applied \cite{Ribeiro05,Ribeiro08}. 
The electrostatic potential is obtained from the polarisation equation by an
FFT Poisson solver with zero-Dirichlet boundary conditions in the (radial)
$x$-direction. Gyrofluid densities are adapted at the $x$-boundaries to ensure
zero vorticity radial boundary conditions for finite ion temperature.  
An Arakawa-Karniadakis scheme is employed for advancing the moment
equations \cite{arakawa66,karniadakis91,Naulin}.

\section{Scaling laws from dimensional analysis} 
\label{analytic}

Blob velocity scalings are commonly deduced from the fluid vorticity
equation. We follow this approach and construct the gyrofluid vorticity
equation to deduce velocity scaling laws. 
The vorticity equation can be obtained upon expressing the gyrocenter ion
density in terms of the electron density and polarisation contribution,  
inserting in the ion gyrocenter density evolution equation and subtracting
the electron density evolution equation \cite{Matthias,Scott07}. 
Up to $\mathcal{O}(b)$ the ion gyrocenter density is
\begin{eqnarray} \label{ion_gyrocenter}
n_{i} = n_\rme - \frac{1}{2} \mu_{i} \nablan^2 p_{i} - \mu_{i} \nablan^2 \phi,
\end{eqnarray}
where the ion pressure is given in terms of the electron particle density $p_{i} = \tau_{i} n_\rme$. The gyroaveraged potential for species $s$ up to $\mathcal{O}(b)$ is
\begin{eqnarray}
\phi_s = \phi +\frac{1}{2}\mu_{i} \tau_{i} \nablan^2 \phi.
\end{eqnarray}
Following Ref.~\cite{Scott07} we obtain
\begin{eqnarray} \label{gf_vorticity}
\mu_{i} \nabla \cdot \frac{\rmd}{\rmd t} \nablan \phi^* = \nablap J_{\spar} -
(1 + \tau_{i}) \mathcal{K} (n_\rme). 
\end{eqnarray}
Here we have introduced the modified potential $\phi^* = \phi + p_{i}$.
The vorticity equation is equivalent to the quasi-neutrality statement of
current continuity, $\nabla \cdot \vek{J} = 0$.  
We identify the divergence of the polarisation current,
\begin{eqnarray}
\nabla \cdot \vek{J}_{\mathrm{pol}} = - \mu_i \nabla \cdot \frac{\rmd}{\rmd t} \nablan \phi^* ,
\end{eqnarray}
and the divergence of the diamagnetic current,
\begin{eqnarray}
\nabla \cdot \vek{J}_{\mathrm{dia}} = -(1 + \tau_i) \mathcal{K} (n_\rme).
\end{eqnarray}
Blob propagation has in a linearisation of the present gyrofluid model been
analytically analysed by Manz \etal in Ref.~\cite{Manz13}. 
Therein the dependence of blob velocity 
on the ion isotope mass is in principle present but not explicitly apparent. 
To clarify, we here restate the calculations of Ref.~\cite{Manz13}, but use the
vorticity eq.~(\ref{gf_vorticity}) with the explicit occurence of $\mu_i$. 
Neglecting parallel currents, employing the blob correspondence
$\partial_x,\partial_y \rightarrow 1 / \sigma$ and $\rmd / \rmd t \rightarrow
i \gamma = i v_b / \sigma$ \cite{Krash_bc}, in terms of the blob width $\sigma$,  
blob velocity $v_b$ and linear growth rate of the instability $\gamma$ and
furthermore identifying $\phi = i v_b \sigma$ (the radial component of the
electric drift), we get (in normalised units):
\begin{eqnarray}
v_b = \frac{1}{\sqrt{2}} \sqrt{\sqrt{f^2 + g^2} - f}, ~\mathrm{where}~ f =
\left( \frac{\tau_i A}{2 \sigma} \right)^2, ~g = \frac{1 + \tau_i}{\mu_i}
\omega_B \sigma A. 
\end{eqnarray}
$A$ is the initial blob amplitude. In the limit of large blobs, $g \gg f$, so that
\begin{eqnarray}
v_b \approx \frac{1}{\sqrt{2}} \sqrt{\frac{1 + \tau_i}{\mu_i} \omega_B \sigma A},
\end{eqnarray}
and for smaller blobs satisfying $g \ll f$ we get
\begin{eqnarray}
v_b \approx \frac{1 + \tau_i}{\tau_i} \frac{\omega_B \sigma^2}{\mu_i}.
\end{eqnarray}
The correspondence with the result of Ref.~\cite{Manz13} is made explicit upon
renormalising, i.e. letting $v_b \rightarrow v_b / \delta c_0, A \rightarrow A
/ \delta, \sigma \rightarrow \sigma / \rho_0, \omega_B \rightarrow 2 L_\perp / R$. 
The limits then are
\begin{eqnarray}
v_b \approx \sqrt{\frac{1 + \tau_i}{\mu_i}} c_0 \sqrt{\frac{2 \sigma }{R} A},
~\mathrm{for} ~\sigma^3 \gg \frac{\mu_i \tau_i^2 A}{4 (1 + \tau_i) \omega_B}, 
\end{eqnarray}
and
\begin{eqnarray}
v_b \approx 2 \frac{c_0}{\mu_i} \frac{1 + \tau_i}{\tau_i} \left(
\frac{\sigma}{\rho_0} \right)^2, ~\mathrm{for} ~\sigma^3 \ll \frac{\mu_i
  \tau_i^2 A}{4 (1 + \tau_i) \omega_B}. 
\end{eqnarray}
For 2-d computations of sufficiently large blobs we consequently expect $v_b
\sim 1 / \sqrt{\mu_i}$, whereas for the 3-d model the expected scaling is not
a priori that clear. In Ref.~\cite{Manz13} 3-d (linear) scaling laws were presented,
where the parallel dynamics was approximated by the Hasegawa-Wakatani closure,
$\nablap J_\spar = \frac{1}{C} \nablap^2 (n_\rme - \phi)$. 

In the following we are going to compare reduced 2-d and full 3-d dynamical
blob simulations for various isotope species with the analytical $1 / \sqrt{\mu_i}$-scaling.

\section{Two-dimensional blob computations} 
\label{two_d}

In this section we numerically analyse the dependence ob filament dynamics on
the normalised ion mass $\mu_i$ by 
reduced 2-d blob simulations of the isothermal gyrofluid eqs.~(\ref{density},\ref{parvel}).
For the computations in this section we us as parameters:
curvature $\omega_B = 0.05$, drift scale $\delta = 0.01$, grid size $L_y = 128
\rho_s$, grid points $N_x = N_y = 256$, initial blob amplitude $A = 1$ and 
Gaussian blob width $\sigma = 10 \rho_s$.  

\begin{figure} 
  \begin{subfigure}{8cm}
    \includegraphics[width=7cm]{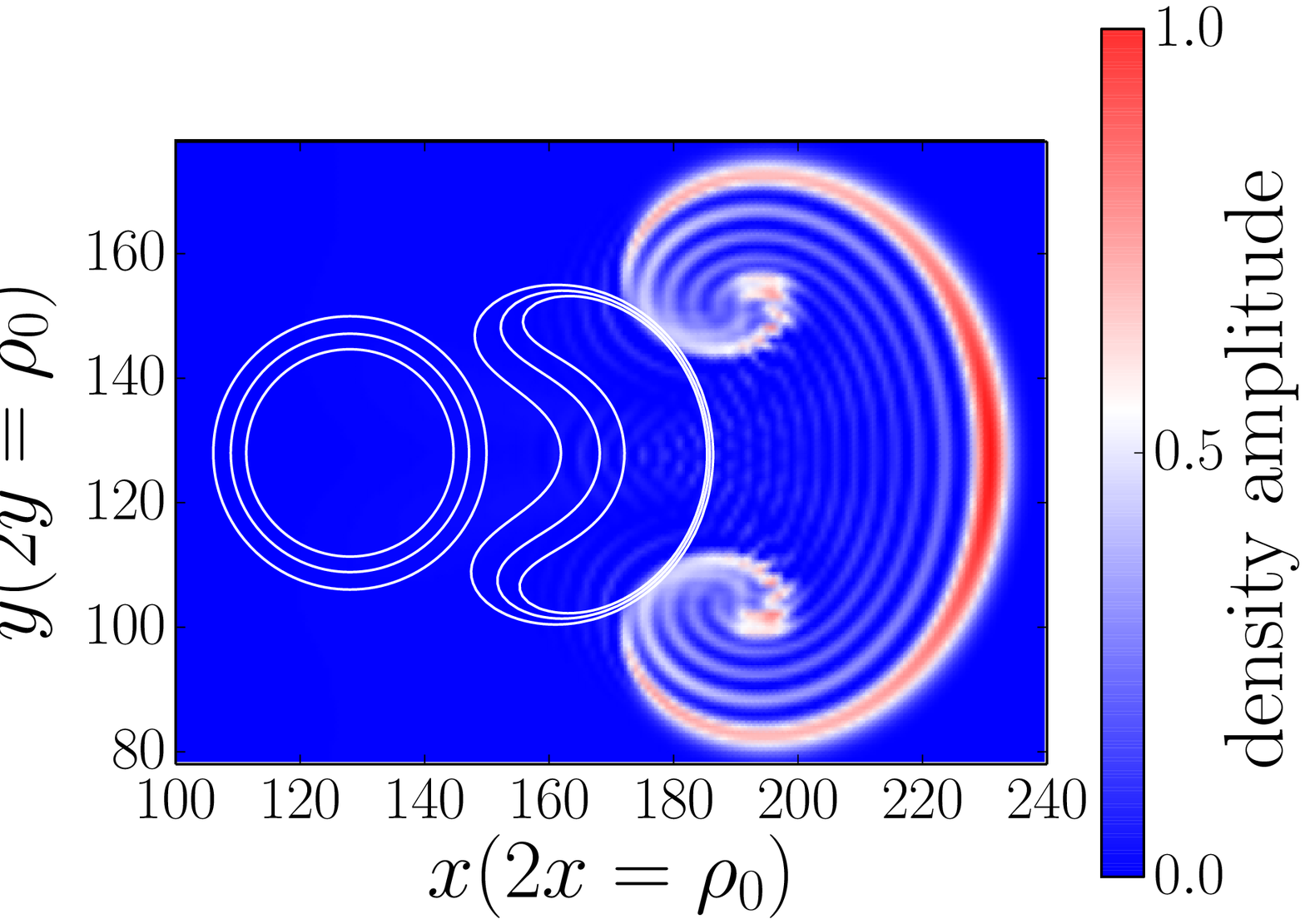} 
  \end{subfigure}
  \begin{subfigure}{8cm}
    \includegraphics[width=7cm]{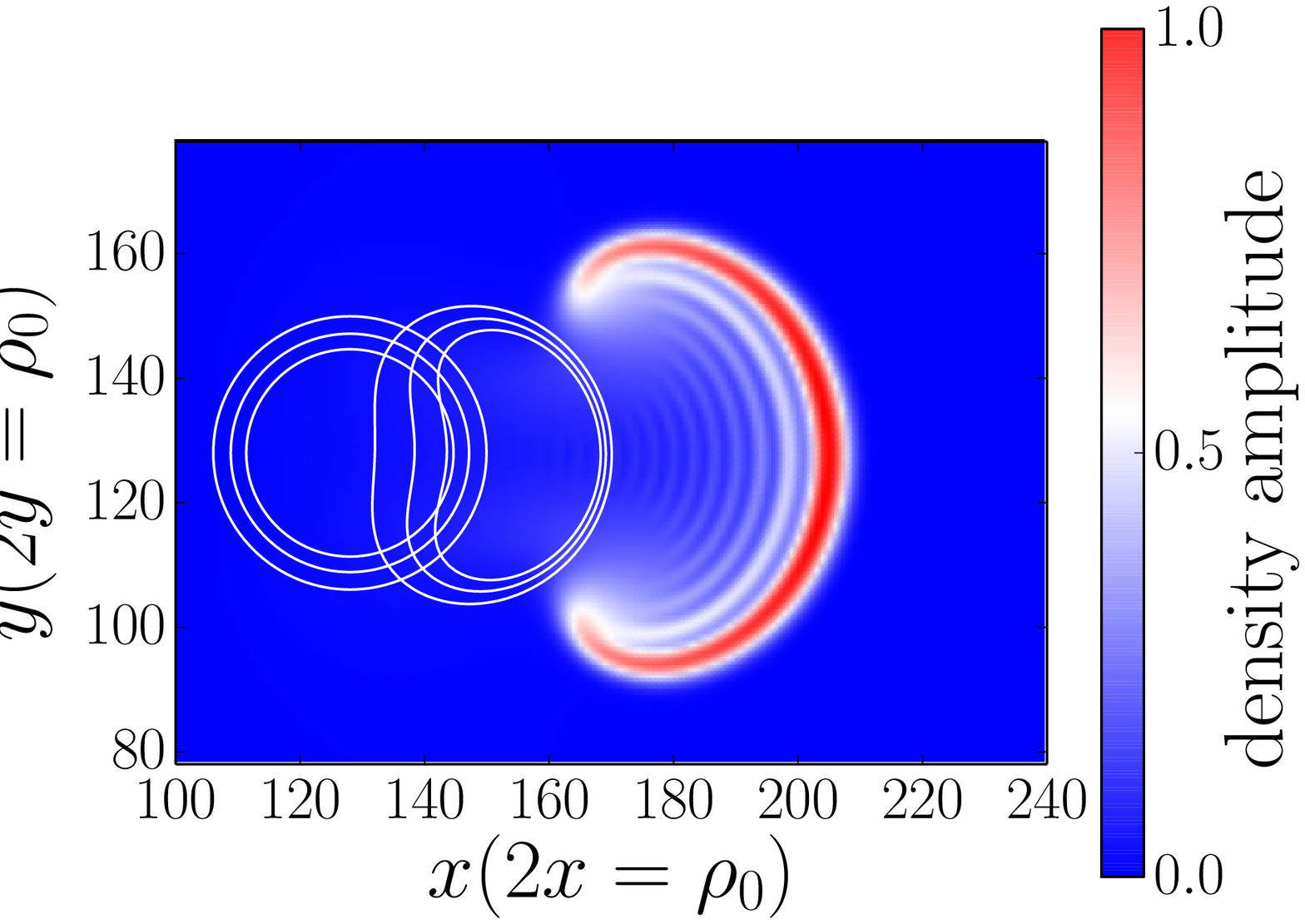} 
  \end{subfigure}  

  \begin{subfigure}{8cm}
    \includegraphics[width=7cm]{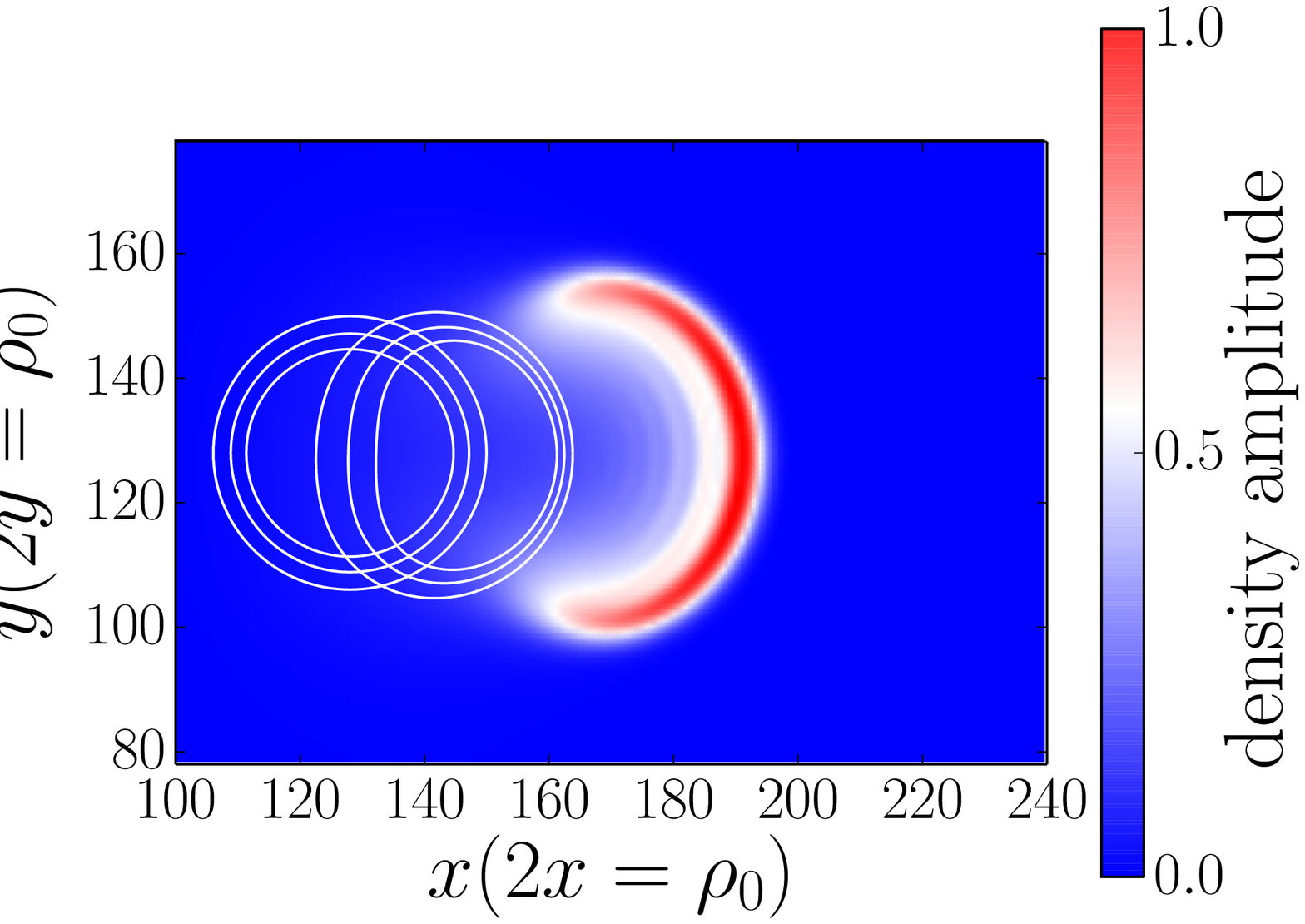} 
  \end{subfigure}
  \begin{subfigure}{8cm}
    \includegraphics[width=7cm]{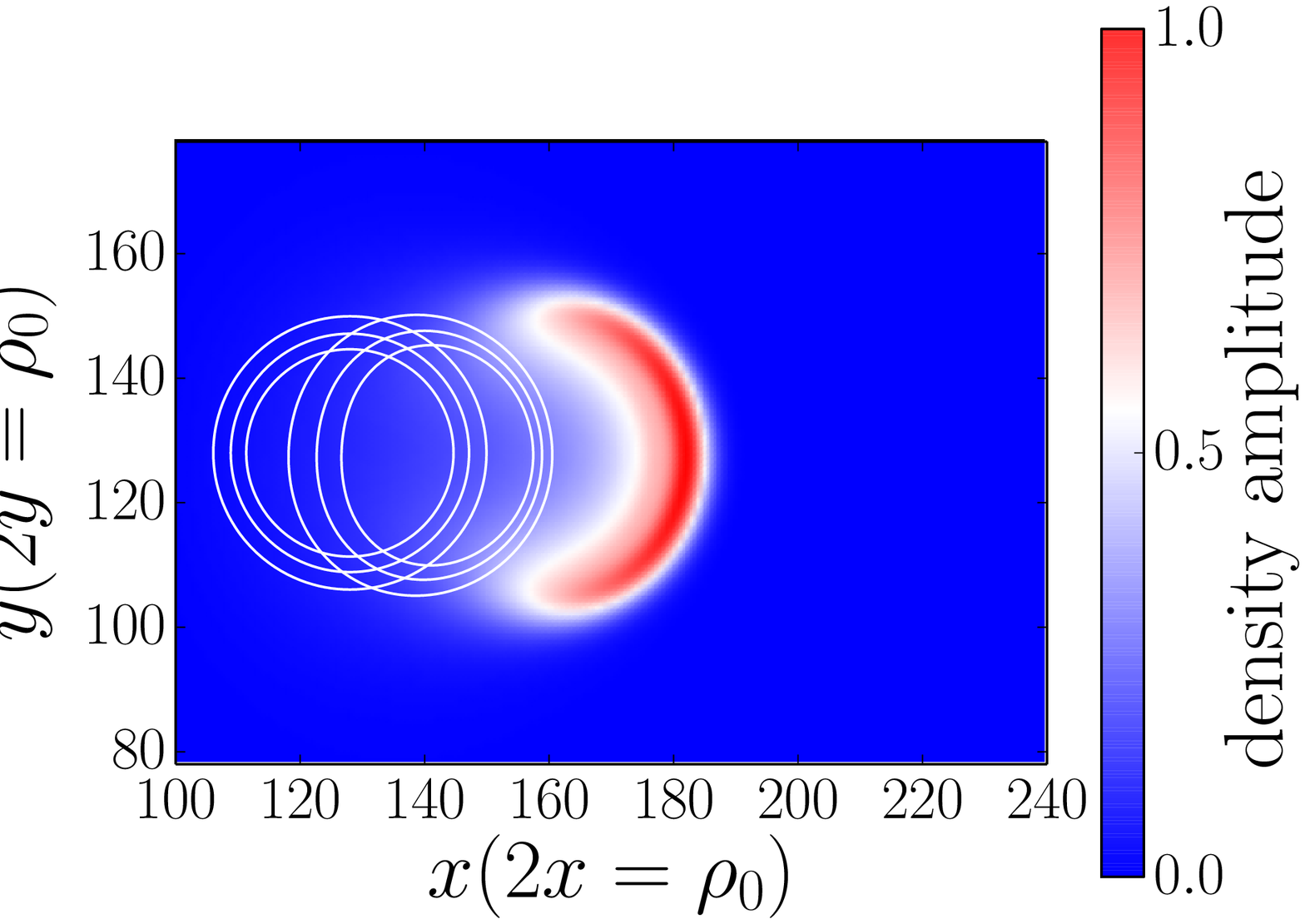} 
  \end{subfigure}
\caption{Electron density contour plots of 2-d cold ion ($\tau_i = 0$) show the
  different blob propagation for several plasma species. White contour lines are snapshots
  at $t=0$, and at $t=3$. The colour plot is drawn at $t=6$.  Isoptopes: protium (top left),
  deuterium (top right), tritium (bottom left) and singly charged helium-4 (bottom right).  }   
\label{2dcontt0}
\end{figure}

Fig.~\ref{2dcontt0} shows contours plots of the electron particle density at
different times of evolution of a seeded blob for several species of cold ions ($\tau_i=0$). 
The initial Gaussian density perturbation $n_e (x,y, t=0) = A \exp \left[ -
  (x^2 + y^2) /  \sigma^2 \right]$ undergoes the familiar transition towards a
mushroom-shaped structure before the blob eventually breaks up due to 
secondary instabilities. This figure illustrates the main point for
the following discussion: lighter isotopes propagate faster than heavier
isotopes. In terms of the (normalising) deuterium mass we consider
$\mu_\mathrm{H} = 1/2$, $\mu_\mathrm{D} = 1$, $\mu_\mathrm{T} = 3/2$ and
$\mu_\mathrm{He+} = 2$ with $\mu_{i} = m_{i} /(Z_i m_{D})$. The species index 
$\mathrm{He+}$ here denotes \textit{singly} charged
helium-4 with $Z_{He+} \equiv 1$. The case of (fully ionised) doubly charged helium-4 isotopes
will be discussed further below in context of 3-d simulations in Sec.~\ref{DeutvHe}.    
Note that the lighter the ion species are, the further the blob is developed
in its radial propagation and evolution at a given snapshot in time.  

For warm ions ($\tau_i >0$) the blob propagation depends on the relative
initialisation of the electron and ion gyrocenter densities.
Commonly, a zero ${\bf E} \times {\bf B}$ vorticity blob initialisation is assumed where 
$n_i (x,y, t=0) = \Gamma_1^{-1} n_\rme (x,y, t=0)$: inserting these into the polarisation
eq.~(\ref{poleq}) results in vorticity $\Omega \equiv \nabla_{\perp}^2 \phi = 0$. This
initialisation for most parameters leads to an FLR induced rapid development of a
perpendicular propagation component in addition to the radial propagation of
the blob, and thus a pronounced up-down asymmetry in $y$ direction.
Alternatively, the electron and ion gyrocenter densities can be chosen as
equal with $n_i (x,y, t=0) = n_e (x,y, t=0)$, so that $\phi \sim n_e$. 
In this case the initial vorticity mostly cancels the FLR asymmetry effect,
and the blob remains more coherent and steady in its radial propagation \cite{Held16}.
The truth may be somewhere in between: as in the experiment blobs are not
``seeded'' (in contrast to common simulations), but appear near the separatrix
from ${\bf E} \times {\bf B}$ drift wave vortices or are sheared off from
poloidal flows, in general some phase-shifted combination of electric
potential and density perturbations will appear. 
For comparison we perform simulations with both of these seeded blob density initialisations.

We note that the $x$ coordinate is effectively pointing radially outwards (in
negative magnetic field gradient direction) at a low-field midplane location
in a tokamak, and the magnetic field here points into the ($x,y$) plane (${\bf e}_z
= {\bf e}_y \times {\bf e}_x$), so that the effective electron diamagnetic
drift direction of poloidal propagation is in the present plots downwards (in
negative $y$ direction).

\begin{figure} 
  \begin{subfigure}{8cm}
    \centering\includegraphics[width=7cm]{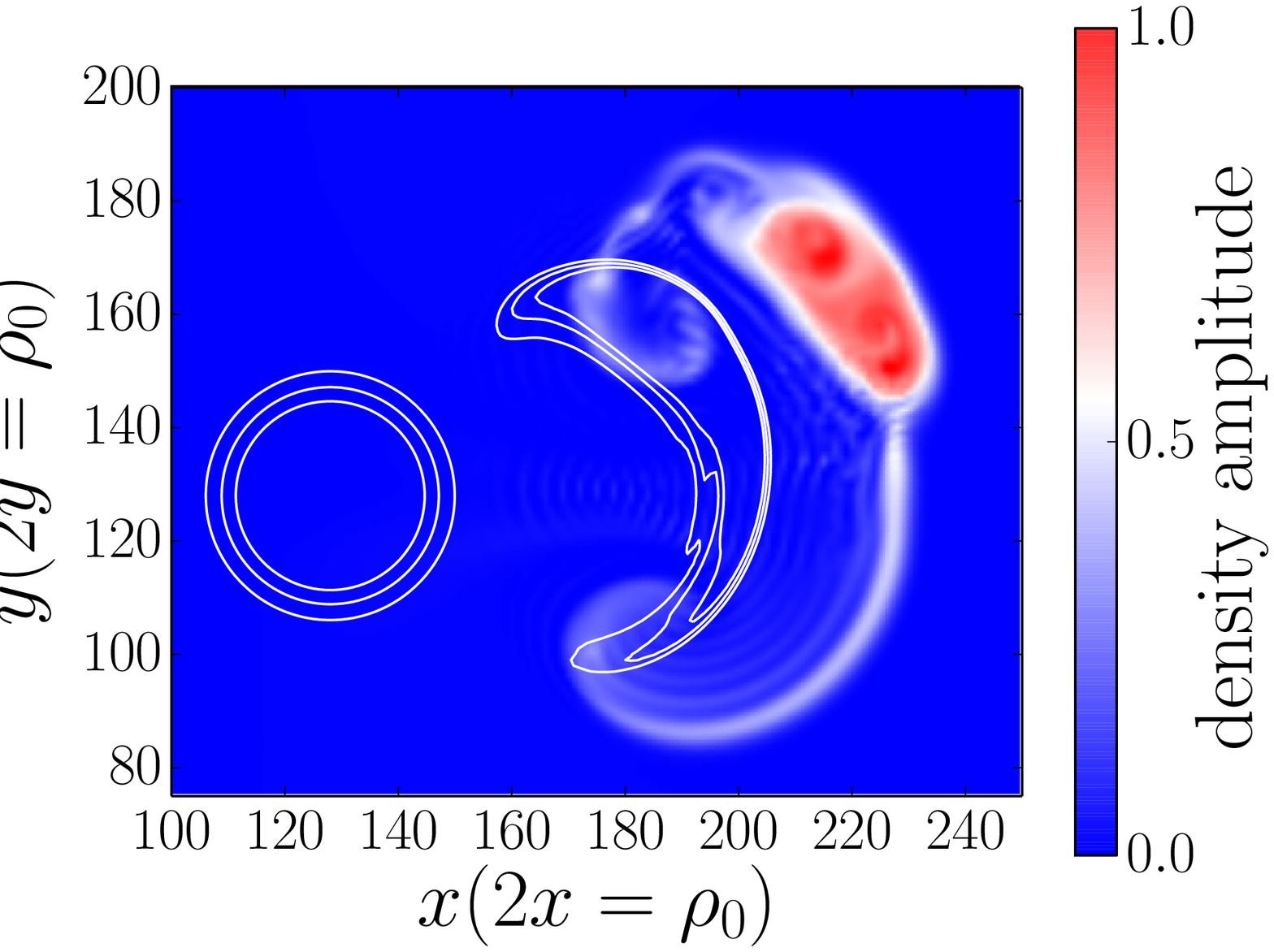} 
  \end{subfigure}
  \begin{subfigure}{8cm}
    \centering\includegraphics[width=7cm]{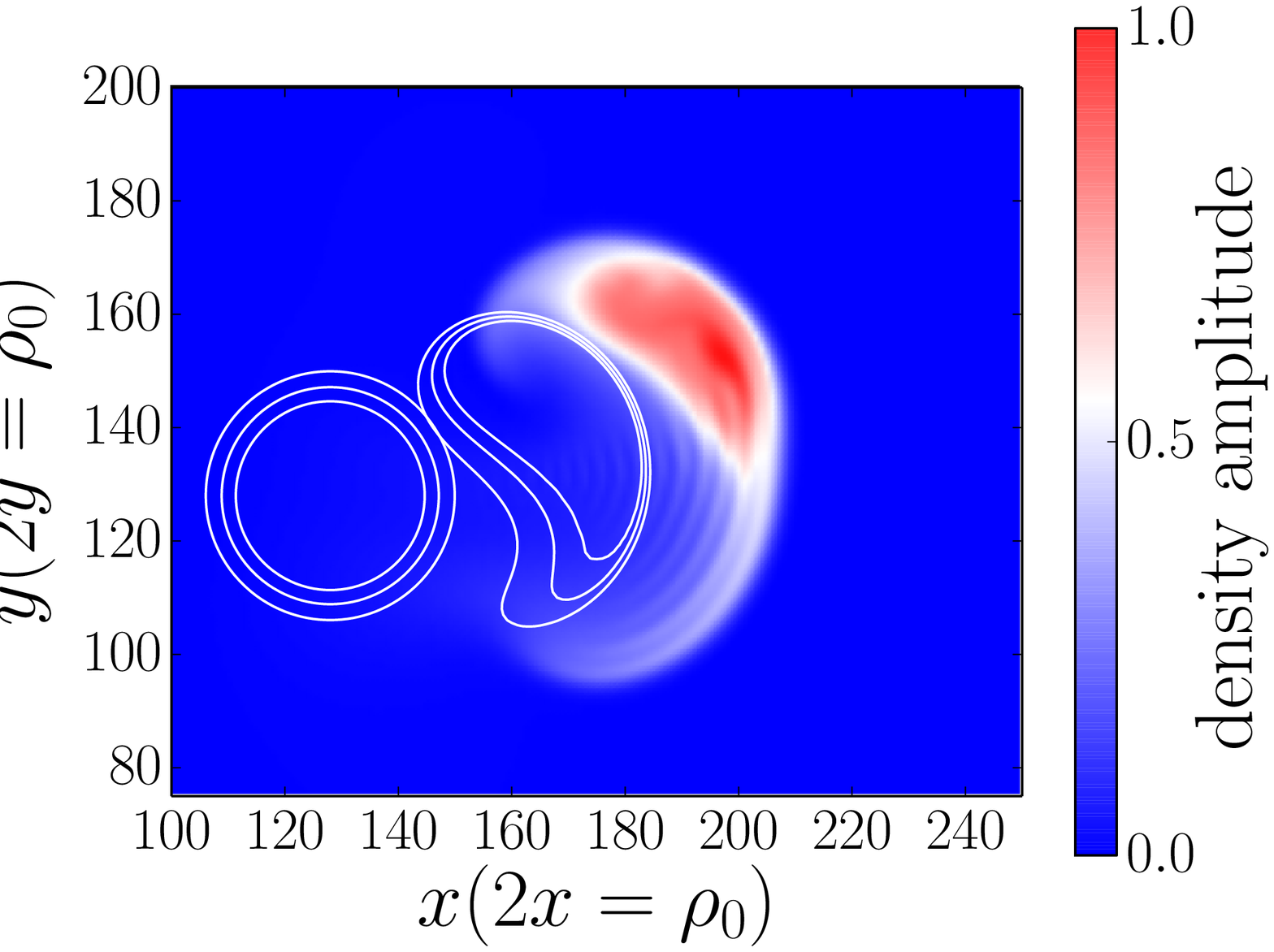} 
  \end{subfigure}
  
  \begin{subfigure}{8cm}
    \centering\includegraphics[width=7cm]{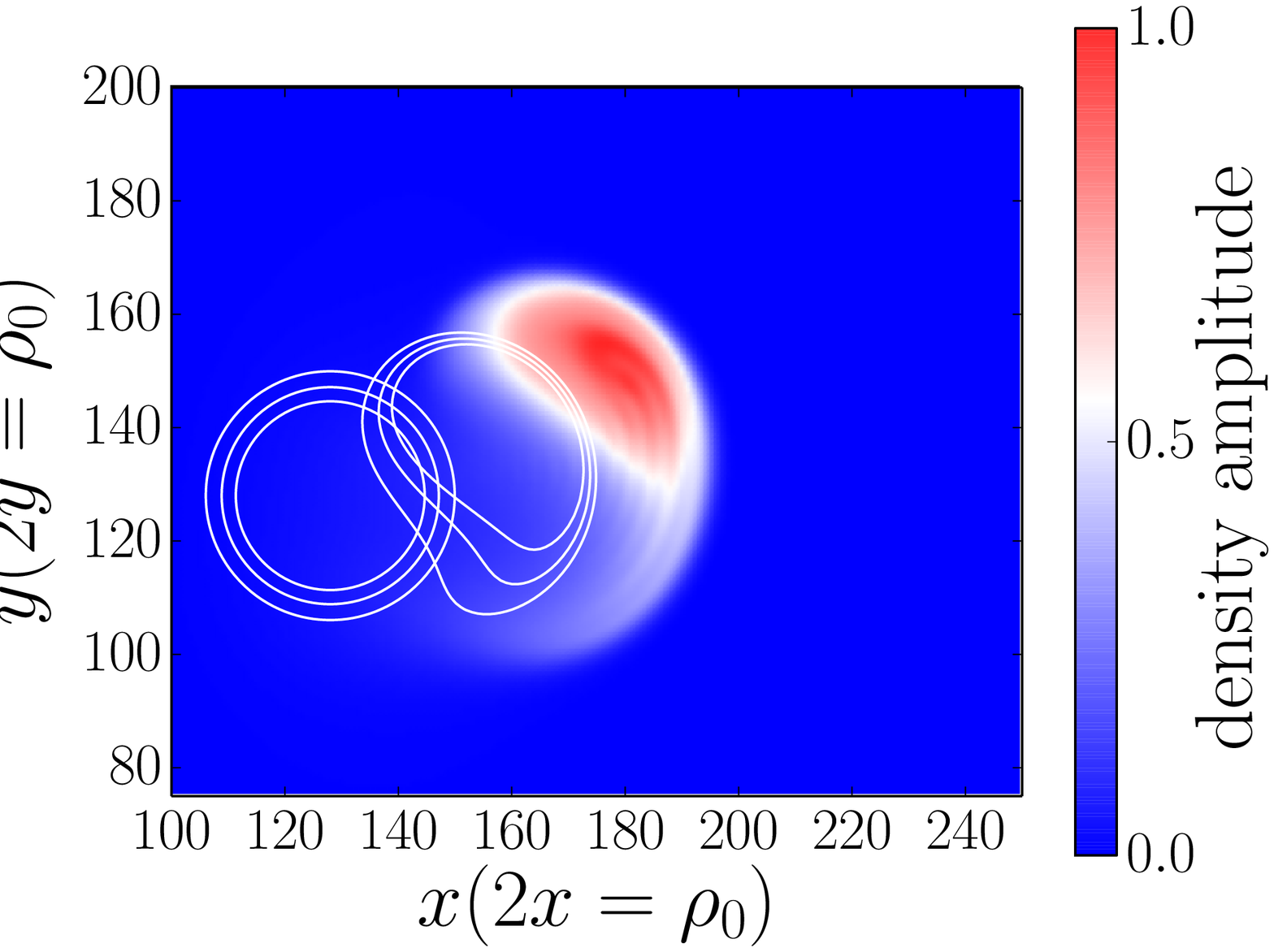} 
  \end{subfigure}
  \begin{subfigure}{8cm}
    \centering\includegraphics[width=7cm]{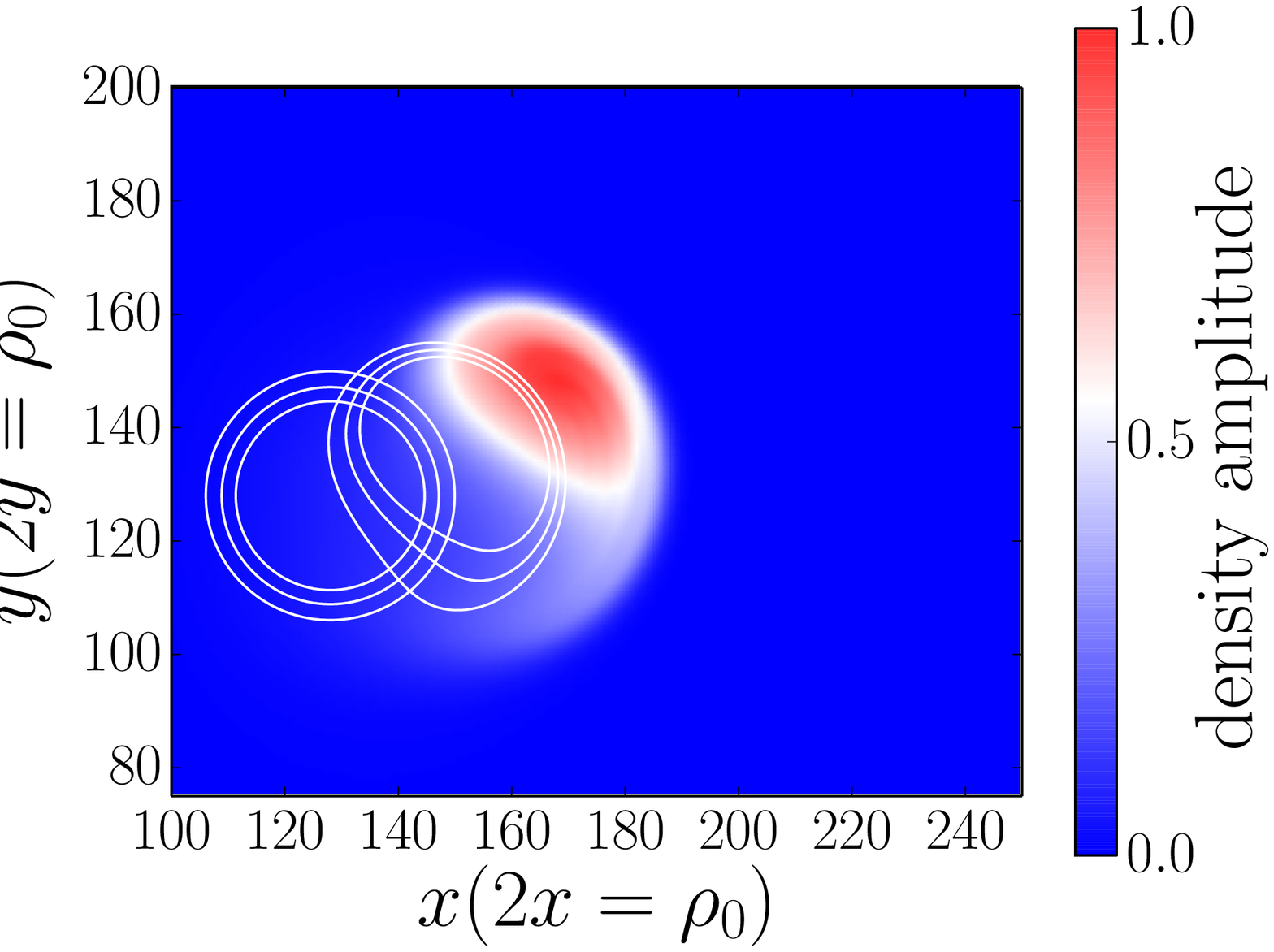} 
  \end{subfigure}
\caption{Electron density contour plots of 2-d warm ion ($\tau_i = 1$) blob
  propagation for different plasma species, using zero-vorticity initial
  conditions. White contour lines are snapshots at $t=0$, and at $t=3$; the
  colour plot is at $t=6$. Protium (top left), deuterium (top right), tritium
  (bottom left) and singly charged helium-4 (bottom right).  }   
\label{2dcontt1}
\end{figure}

\begin{figure} 
  \begin{subfigure}{8cm}
    \centering\includegraphics[width=7cm]{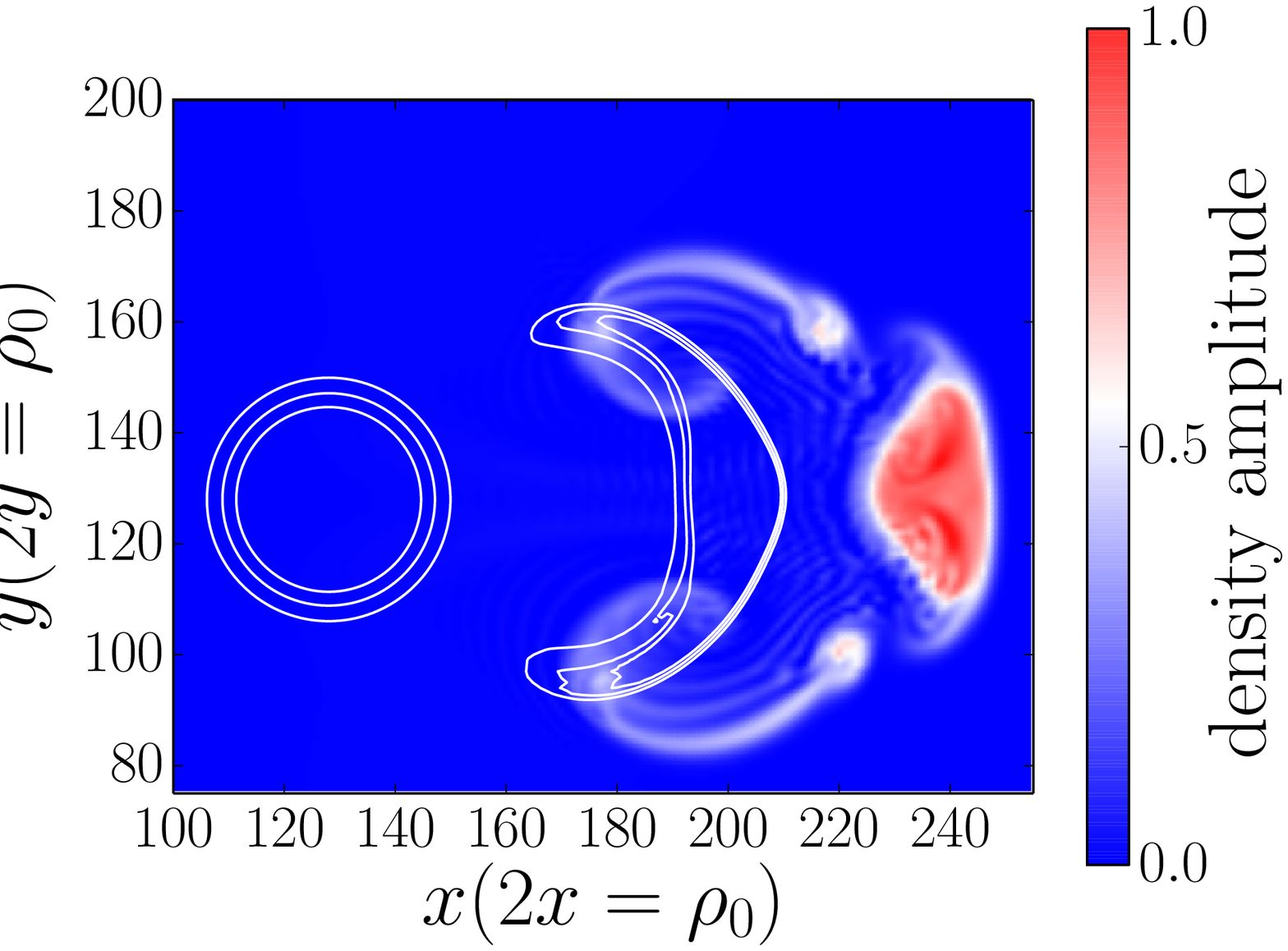} 
  \end{subfigure}
  \begin{subfigure}{8cm}
    \centering\includegraphics[width=7cm]{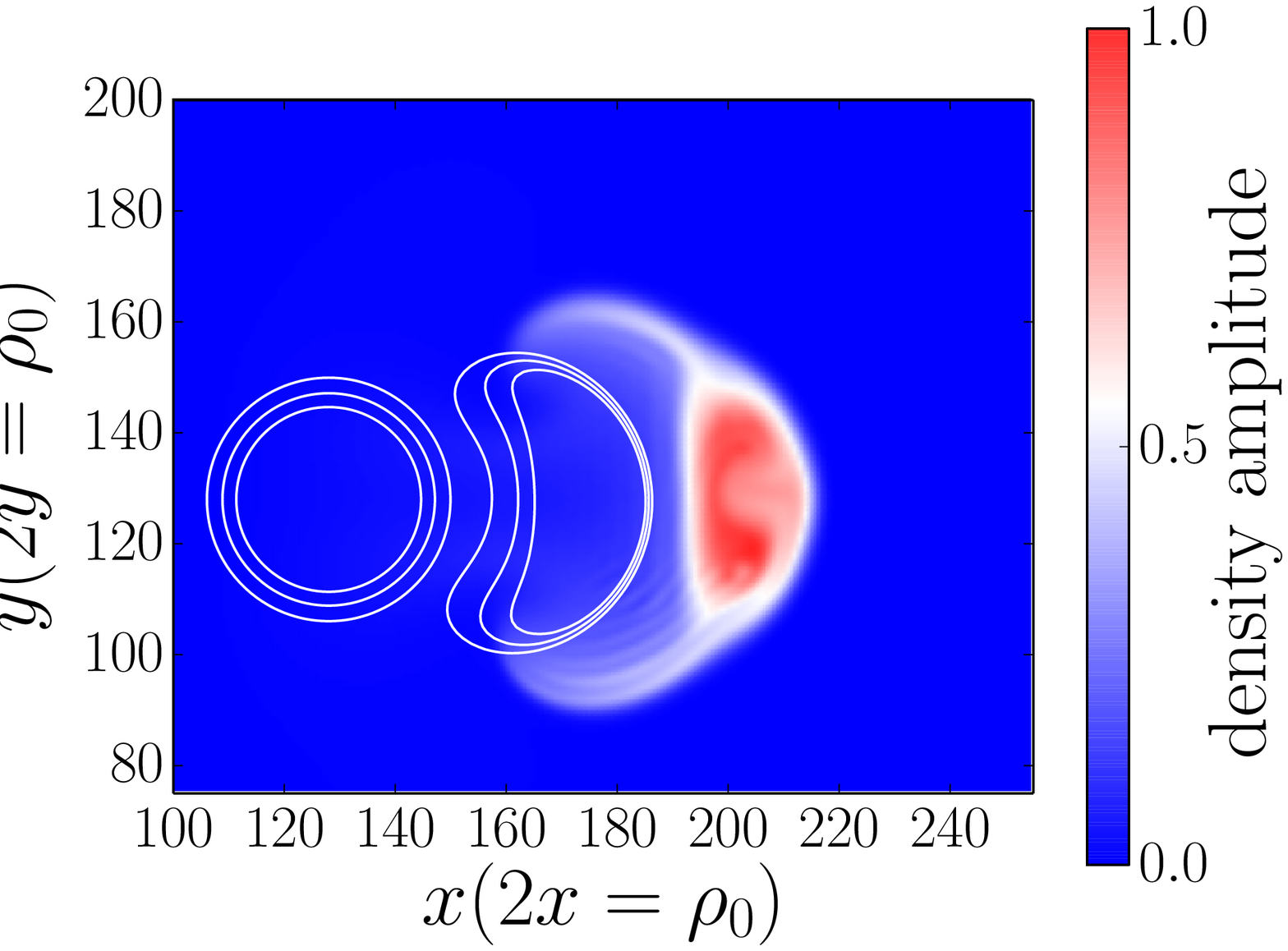} 
  \end{subfigure}
  
  \begin{subfigure}{8cm}
    \centering\includegraphics[width=7cm]{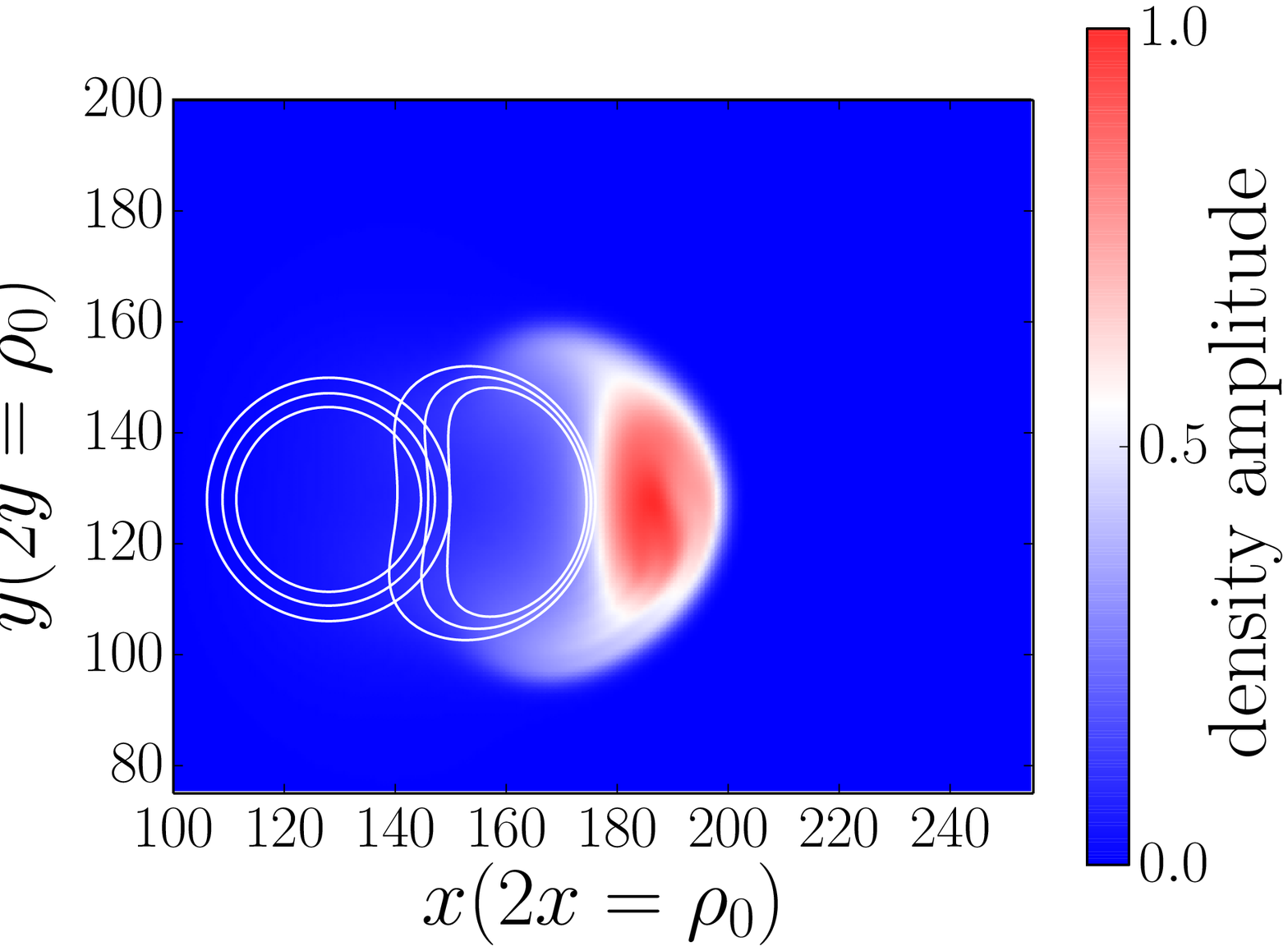} 
  \end{subfigure}
  \begin{subfigure}{8cm}
    \centering\includegraphics[width=7cm]{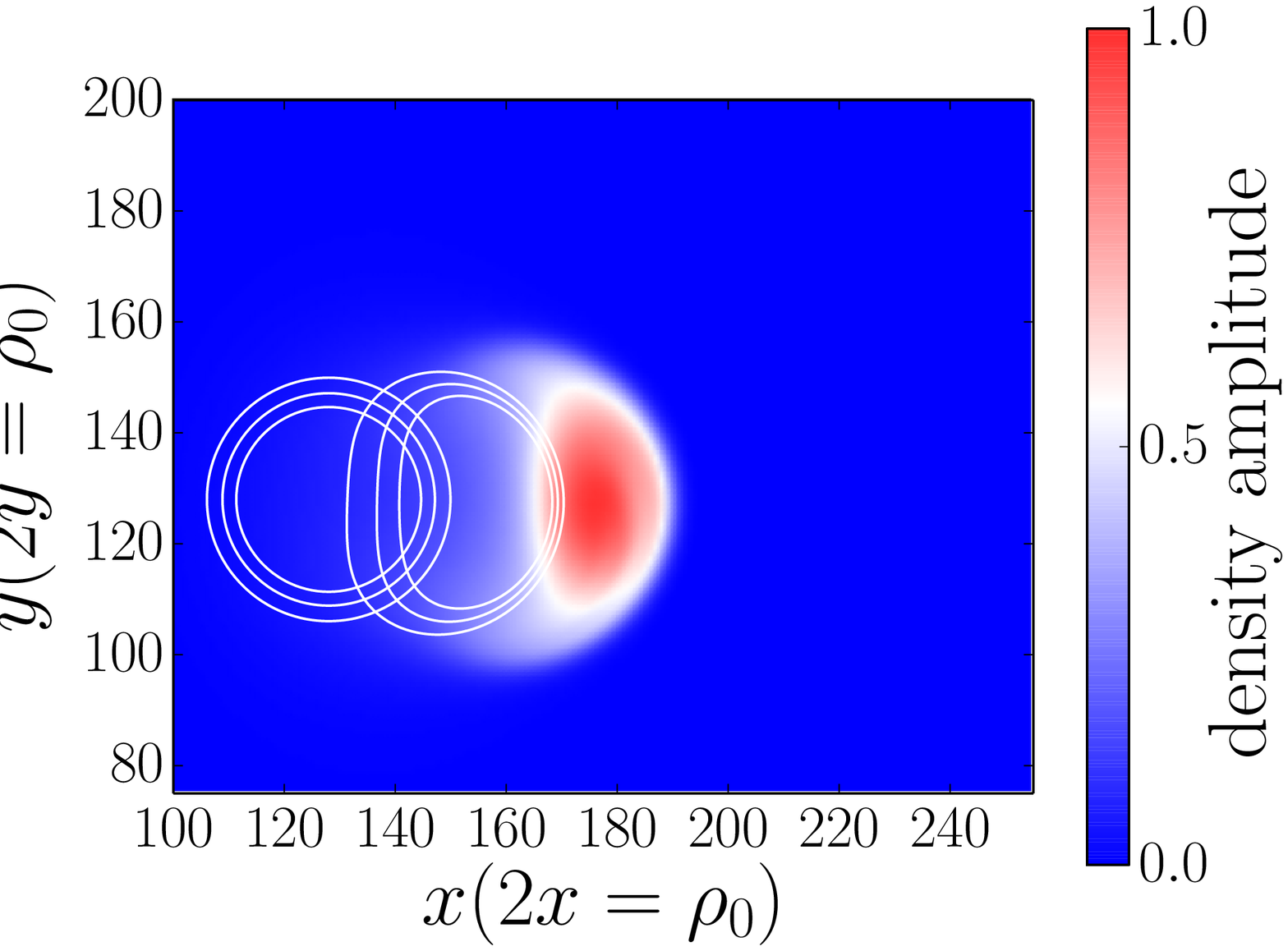} 
  \end{subfigure}
\caption{Electron density contour plots of 2-d warm ion ($\tau_i = 1$) blob
  propagation for different plasma species, using non-zero vorticity initial
  conditions. White contour lines are snapshots at $t=0$, and at $t=3$; the
  colour plot is at $t=6$. Protium (top left), deuterium (top right), tritium
  (bottom left) and singly charged helium-4 (bottom right).  }   
\label{2dcontt1ng}
\end{figure}

Fig.~\ref{2dcontt1} shows blob propagation for warm ions with $\tau_i = 1$,
initialised with the zero vorticity condition. 
For comparison, we present in Fig.~\ref{2dcontt1ng} the propagation for the same
parameters but initialised with equal electron and ion gyrocenter densities. 
Clearly, the latter cases with initial non-zero vorticity  $\Omega = \nabla_{\perp}^2
\phi$ results in faster and more coherent radial propagation, whereas the
zero vorticity cases exhibit significant poloidal translation through the FLR
induced spin-up.
Regardless of initialisation, blobs of light ion species with small $\mu_i$
travel faster and are further developed at a given time compared to heavier species.

\begin{figure} 
  \begin{subfigure}{8cm}
    \centering{\includegraphics[width=7cm]{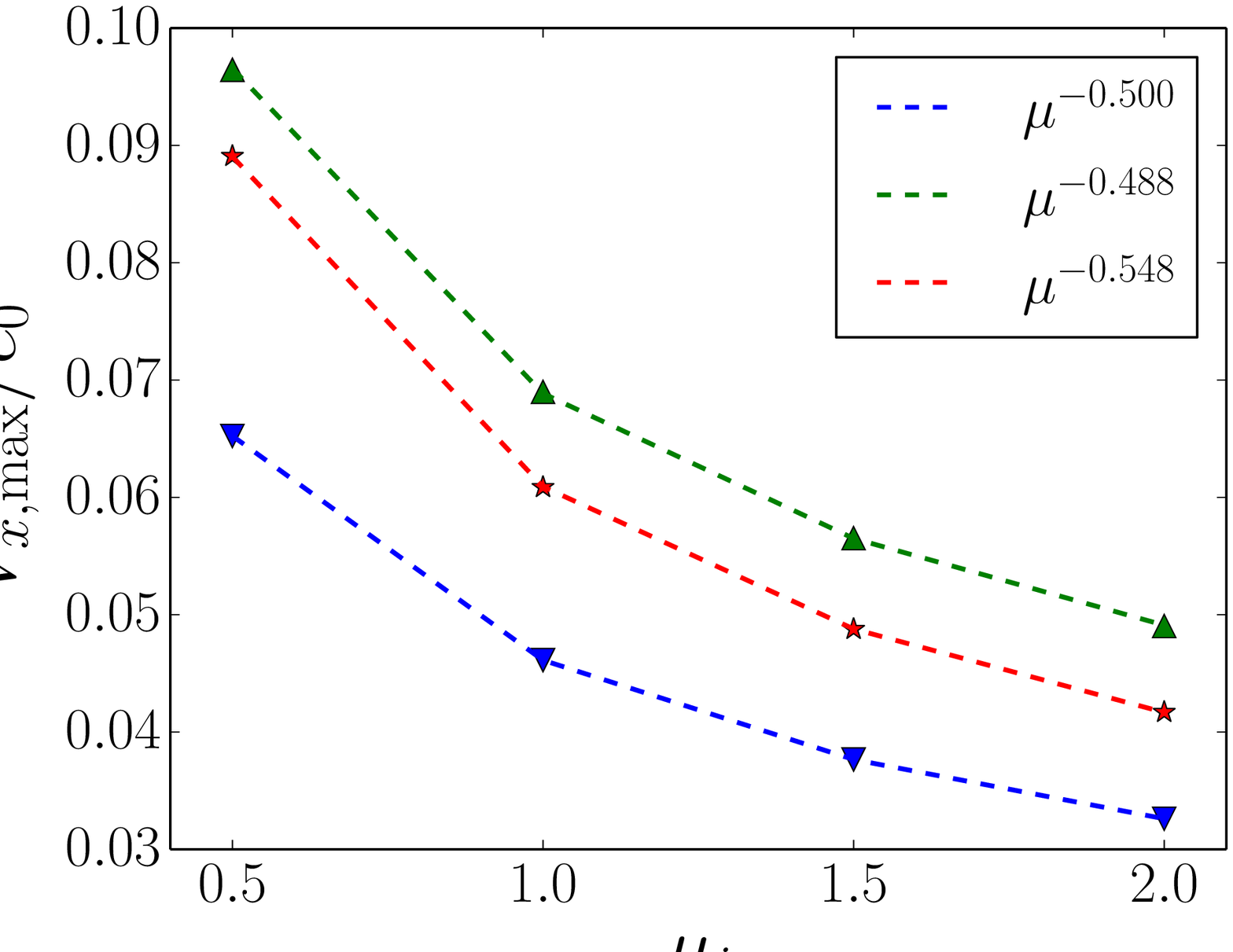}} 
  \end{subfigure}
  \begin{subfigure}{8cm}  
    \centering{\includegraphics[width=7cm]{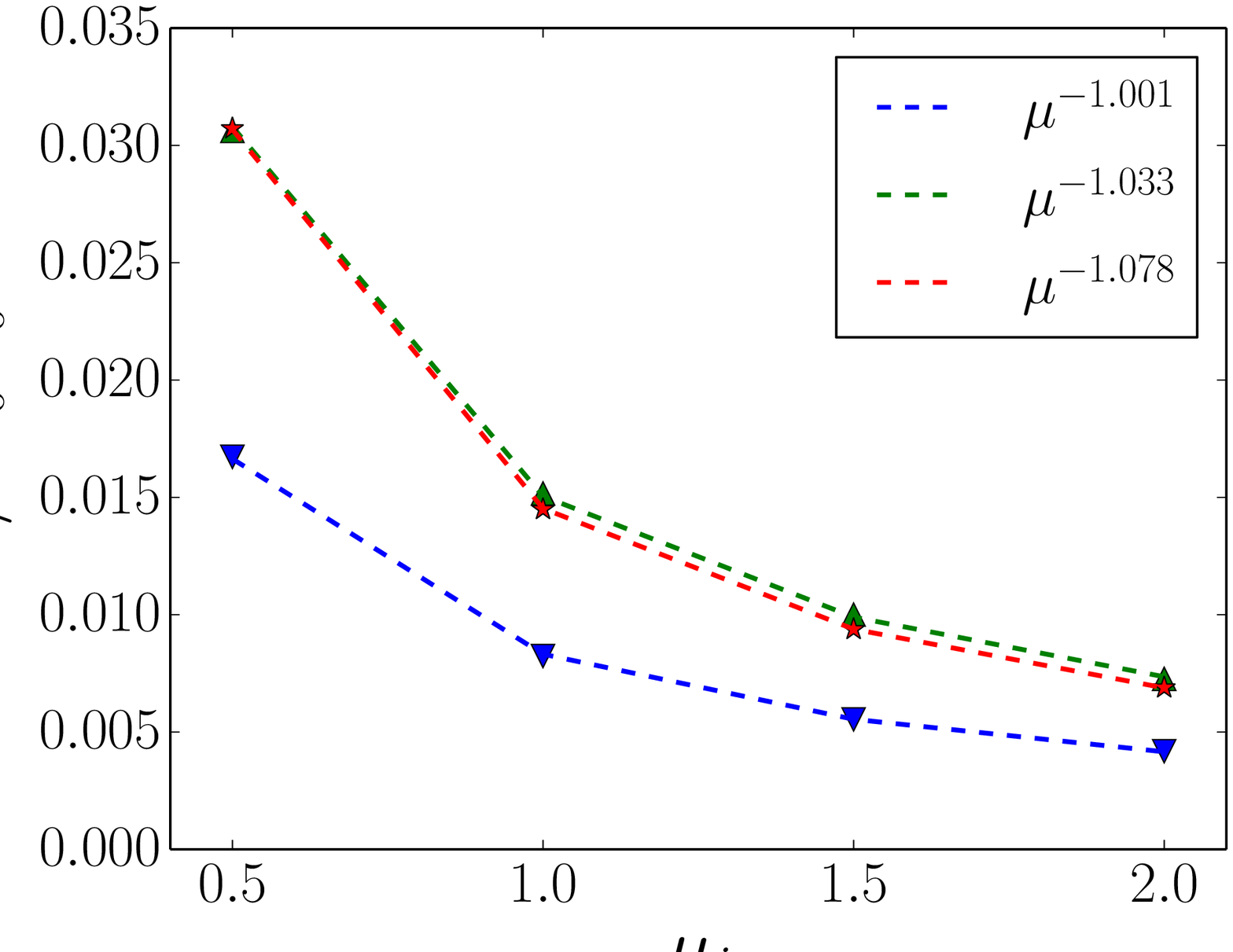}}
  \end{subfigure}
\caption{Maximum radial 2-d center-of-mass velocity (left) and average radial
  acceleration (right) for cold ions (blue), warm ions initialised with zero
  vorticity (red), and warm ions initialised with non-zero vorticity (green).}   
\label{2dvels}
\end{figure}

Relevant quantities which determine the intermittent blob related transport properties
of the tokamak SOL are the maximum blob velocity and acceleration.
In Fig.~\ref{2dvels} we present maximum radial center-of-mass velocities
$V_{x, \mathrm{max}}$ and the average radial acceleration $A$ as a function of
the the ion mass parameter $\mu_i$.
The different symbols/colours represent cases with  cold ($\tau_i = 0$, blue
lower curves) and warm  ($\tau_i = 1$) ions, with both types of initial
conditions used on the latter: the zero vorticity condition is depicted in red
(middle curves) and the $n_e=n_i$  condition in green (upper curves).

It can be seen that the maximum radial blob velocity is slightly larger for $n_e=n_i$
initialisation due to the mainly radial propagation (left figure), but the
average acceleration is for both cases nearly equal (right figure).

The radial center-of-mass position is given by $X_{\mathrm{c}} = [ \int \rmd x \rmd y ~
  x n_\rme ] / [ \int \rmd x \rmd y ~ n_\rme ]$.
Taking the temporal derivative gives the radial center-of-mass
velocity, $V_x = \rmd X_\mathrm{c} / \rmd t$. 
The maximum of $V_x (t)$, $V_{x, \mathrm{max}} = \mathrm{max} \{ V_x (t) \}$
and the corresponding time for the occurence of the maximum, $t_\mathrm{max}$,
then give a measure of the average radial acceleration, 
$A = V_{x, \mathrm{max}} / t_\mathrm{max}$.

Clearly, an inverse dependence of velocities and acceleration on effective
ion mass $\mu_i$ can be inferred for all cases. 
For cold ions, the only mass dependence in the present 2-d isothermal
gyrofluid model, lies in the gyrofluid polarisation equation, carrying over
the mass dependence of the polarisation drift in a fluid model. 
As deduced from the basic linear considerations in Sec.~\ref{analytic}, the
maximum blob velocity scales inversely with the square root of the ion species
or isotope mass: the plotted fits are close to the expected lines $V_{x,
  \mathrm{max}} \sim \mu^{-0.5}$.

From dimensional analysis it follows that the acceleration should scale
according to $A \sim \gamma^2 \sigma$, where $\gamma$ is the growth-rate 
of the linear instability. For $\gamma \sim 1 / \sqrt{\mu_i}$, we expect 
$A \sim 1 / \mu_i$ for cold ions: this is confirmed in Fig.~\ref{2dvels}
(right) where the fitted exponents are close to $-1$. 
Warm ion simulations also feature $\mu_i$ species mass dependence through the FLR
  operators $\Gamma_0(b)$ and $\Gamma_1(b)$, where $b = \rho_i^2 k_{\perp}^2 =
  \mu_i \tau_i \rho_0 k_{\perp}^2$.

We find that for the parameters at hand, the maximum radial velocity for warm
ions with zero vorticity initialisation is higher compared to cold ions, with
a slightly increased isotopic dependence (seen in an exponent $-0.548$
compared to $-0.500$). 

Initialising with non-zero vorticity yields approximately $50 \%$ increased
velocities compared to cold ions, and slightly weakens the isotopic dependence
(expressed by an exponent $-0.488$). This can be attributed to the
mass dependence in the FLR operators, which we further discuss below in Sec.~\ref{nozGF}.

\section{Three-dimensional filament computations}
\label{three_d}

In three dimensions, when the blob extends into an elongated filament along
the magnetic field lines, additional physics enters into the model. 
The basic picture of interchange driving of filaments by charging through $\nabla
B$ and curvature drifts to produce a net outward $\vek{E} \times \vek{B}$ 
propagation still remains valid. However, the total current continuity balance
now also involves parallel currents: 
$ - \nabla \cdot \vek{J}_\mathrm{pol} = \nablap J_\spar + \nabla \cdot \vek{J}_\mathrm{dia}$. 

The detailed balance among the current terms determines the overall motion of
the filament. Furthermore, blob filaments in the edge of toroidal magnetised
plasmas generally tend to exhibit ballooning in the unfavourable curvature
region along the magnetic field.  
The parallel gradients in a ballooned blob structure also lead to a parallel
Boltzmann response, mediated mainly through the resistive coupling of $\nablap (\phi -
n_\rme)$ to $C J_{||}$ in eq.~(\ref{parvel}).  
This tends towards (more or less phase shifted) alignment between the electric
potential and the perturbed density, which strongly depends on the
collisionality parameter $C$.  

For low collisionality, the electric potential in the blob evolves towards
establishment of a Boltzmann relation in phase with the electron density along $\vek{B}$, so
that $n_\rme \sim \exp(- \phi) \sim \phi $. This leads to reduced
radial particle transport, and the resulting spatial alignment of the
potential with the blob density perturbation produces a rotating vortex along
contours of constant density, the so-called Boltzmann spinning \cite{angus12,angus14}. 
Large collisionality leads to a delay in the build-up of the potential within
the blob, so that the radial interchange driving can compete with the parallel
evolution, and the perpendicular propagation is similar to the 2-d scenario.

In the following we investigate how 3-d filament dynamics is depending on the
ion mass. Clearly, we expect an impact in addition to the 2-d effects found in
the previous section, since (i) the parallel ion velocity is inversely
dependent on ion mass (but is for any ion species slow compared to the
electron velicity), (ii) the sheath boundary coupling constants are mass
dependent, and (iii) the basic dependence on the ion mass in the polarisation
current will play a more role complicated role compared to the 2-d model.  

For our present study we chose the free computational parameters
basically identical to the 2-d case above: drift scale $\delta = 0.01$,
curvature $\omega_B=0.05$, blob amplitude $A=1$ and perpendicular blob width $\sigma=10$.
The Gaussian width of initial parallel density perturbation is given by
$\Delta_z=\sqrt{32}$, which represents a slight ballooning with some initial
sheath connection: 
\begin{equation}
n_\rme^{3D} (t = 0, x, y, z) =  n_{e \perp}  \cdot \exp
\left[\frac{-(z - z_0)^2}{\Delta^2} \right] 
\end{equation}
where $z_0$ is the parallel reference coordinate at the outboard mid-plane and
$n_{e \perp}(x,y)$ is the perpendicular Gaussian initial perturbation
introduced in Sec.~\ref{two_d}.  
In this section we first focus on zero vorticity initial conditions, non-zero
conditions will be discussed further below.

The perpendicular domain size is $L_x = L_y = 128 \rho_s$ with a grid
resolution of $N_x = N_y = 256$. The number of parallel grid points is varied
between $N_z=8$ and $16$. The filament simulations have been tested for convergence
with respect to the number of drift planes up to $N_z =32$: $N_z = 8$ yields
qualitatively and quantitatively similar results to $N_z = 16$. 
The plots showing colour cross sections throughout this article are taken from
simulations with $N_z = 8$, and the presented quantitative results have been
obtained for $N_z = 16$.  

We here set $\wh{\beta} = 0$ as electromagnetic effects in the SOL are thought to
be of minor importance for the present discussion \cite{Manz13}. 
The collisionality parameter is chosen in $C = 0.5 - 100$ to cover a
likely range of tokamak SOL values. 
Typical values for the collisionality parameter for the SOL in ASDEX Upgrade
L-mode plasmas have in the literature \cite{Manz13} been reported as $C \sim 1
- 100$, and a reference characteristic collisionality in Ref.~\cite{Easy1} 
for MAST has been given as $C \sim 2$. 
\begin{figure} 
  \begin{subfigure}{8cm}
    \centering{\includegraphics[width=7cm]{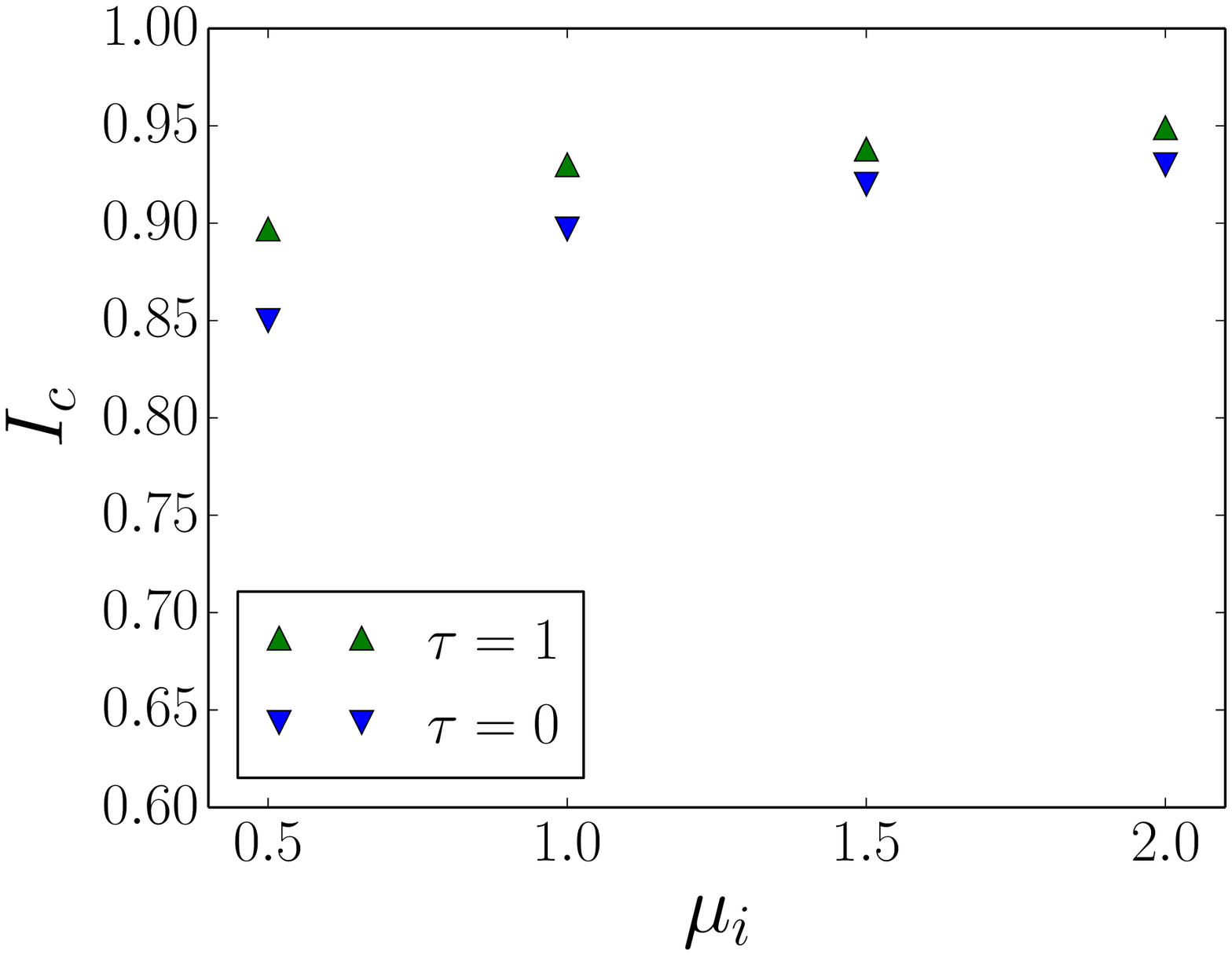}} 
  \end{subfigure}
  \begin{subfigure}{8cm}  
    \centering{\includegraphics[width=7cm]{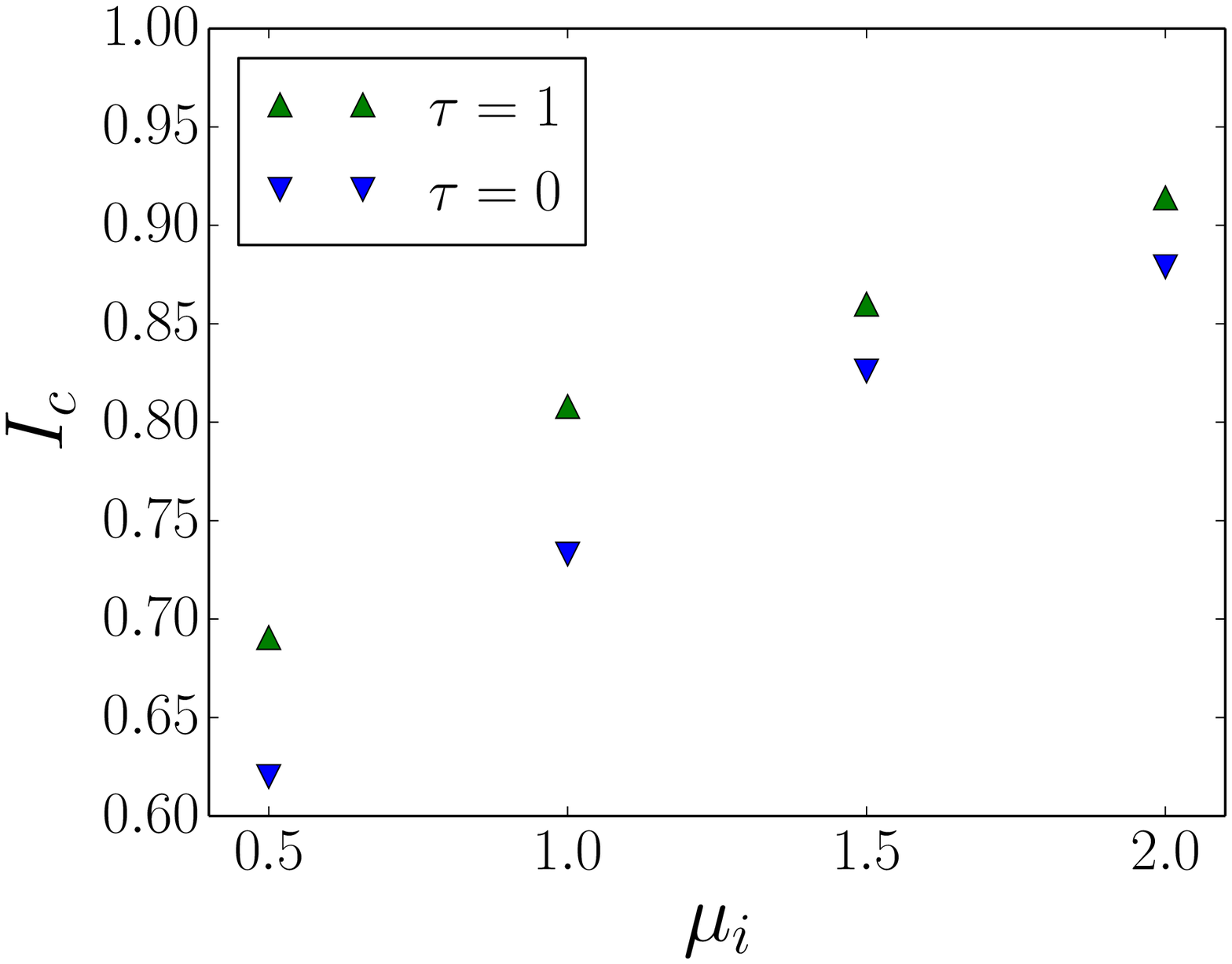}}
  \end{subfigure}
\caption{Cross section of 3-d filament at outboard-midplane location ($z =
  z_0$): electron density contour plot at $t = 6$ for warm deuterium ions
  ($\tau_i = 1$). Boltzmann spinning is dominant for intermediate
  collisionality parameter ($C = 10$, left), and reduced for increased
  collisionality ($C = 100$, right).}   
\label{3dcont_Boltz}
\end{figure}

Fig.~\ref{3dcont_Boltz} illustrates the dependence of filament evolution
with respect to collisionality dependent Boltzmann spinning for warm deuterium ions. 
The case with $C = 10$ represents the strong Boltzmann spinning (drift wave)
regime, where density and potential perturbations are closely aligned and the
radial filament motion is strongly impeded. 
Increasing the collisionality to $C = 100$ reduces parallel electron dynamics
and so effectively increases the lag of potential build-up within the density
blob perturbations, so that the Boltzmann spinning is reduced. 
The substantial perpendicular motion component is in this case partly caused
by FLR effects like in the corresponding 2-d case for $\tau_i=1$.
The $\tau_i$ contribution to the ion diamagnetic curvature term results in 
enhanced radial driving of the blob compared to cold ion cases.

\begin{figure} 
  \begin{subfigure}{8cm}
    \centering{\includegraphics[width=7cm]{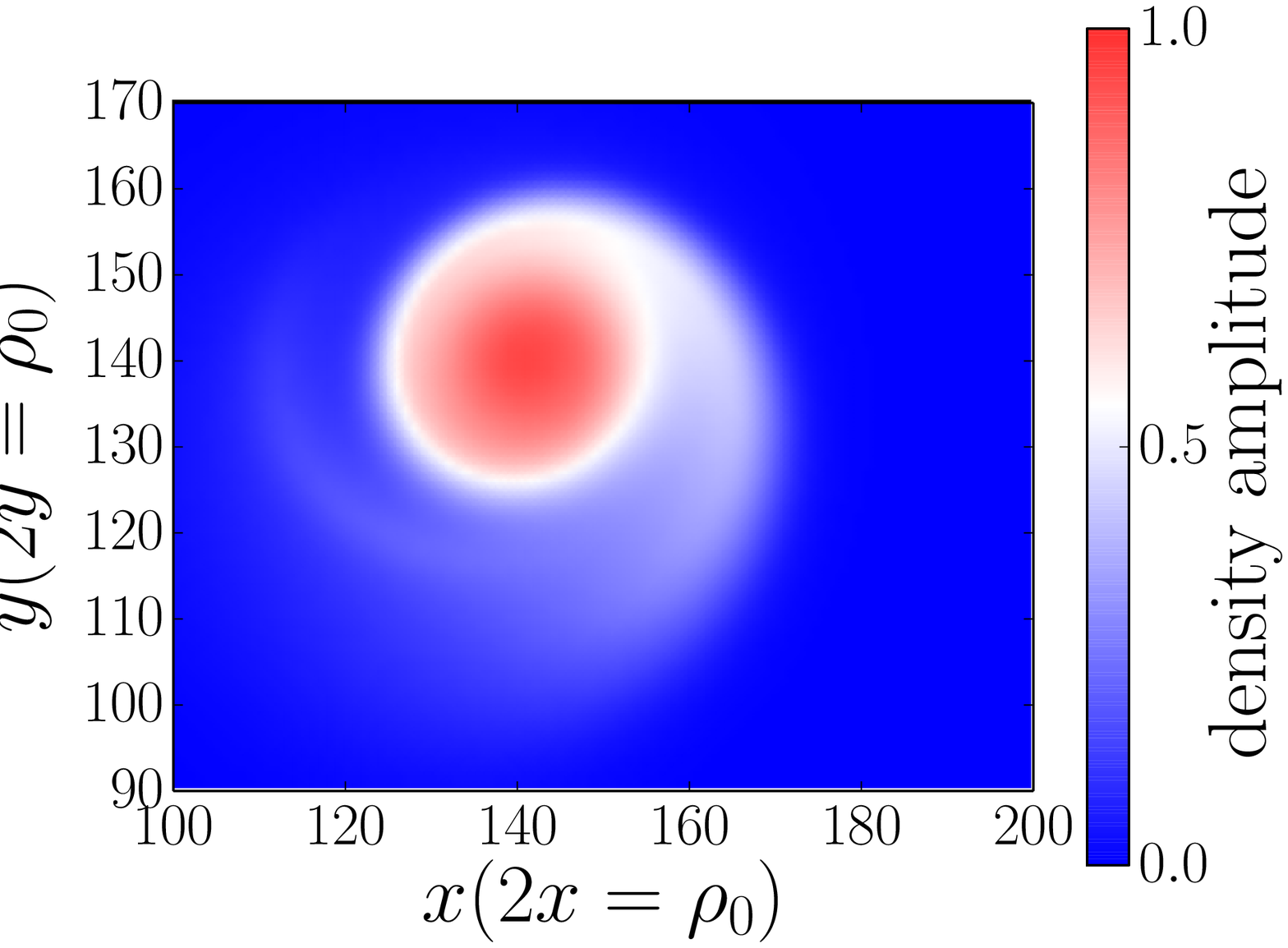}} 
  \end{subfigure}
  \begin{subfigure}{8cm}  
    \centering{\includegraphics[width=7cm]{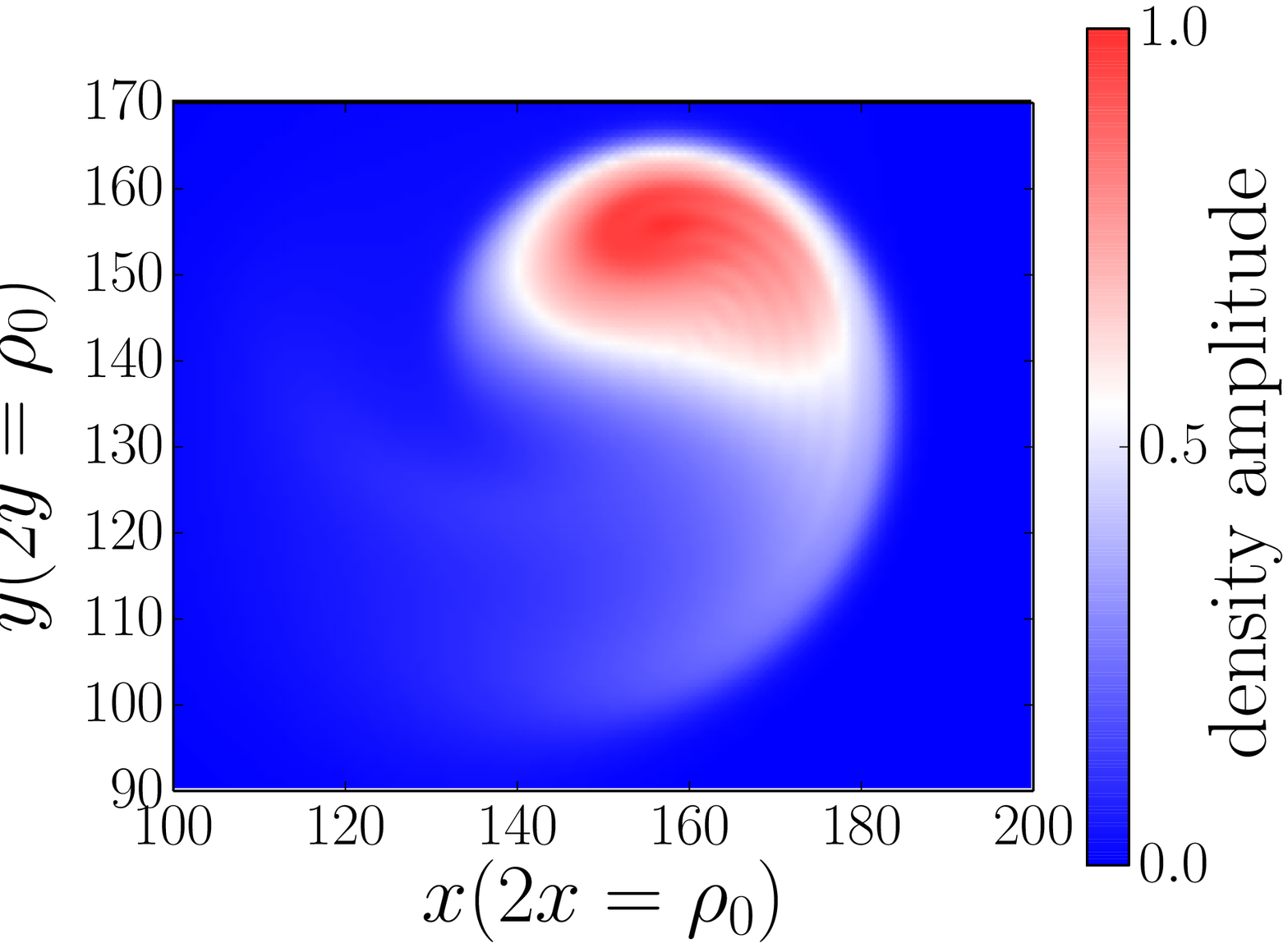}}
  \end{subfigure}
\caption{Blob compactness at $t = 5$ analysed at the outboard midplane ($z =
  z_0$): strong Boltzmann spinning ($C = 10$, left) and weak spinning ($C = 100$, right).}   
\label{3dcompactn}
\end{figure}

A measure for blob compactness can be introduced \cite{Madsen} by
\begin{eqnarray}
I_c (t) = \frac{\int \rmd x \int \rmd y n_\rme (x,y,t) h(x,y,t)}{\int \rmd x
  \int \rmd y n_\rme (x,y,t=0) h(x,y,t=0)}, 
\end{eqnarray}
where the Heavyside function $h(x,y,t)$ is defined as
\begin{eqnarray}
h (x,y,t) = 1 ~ \quad \mathrm{if} ~ (x - x_\mathrm{max} (t))^2 + (y -
y_\mathrm{max} (t))^2 < \sigma^2, 
\end{eqnarray}
and zero elsewhere. That is, the integral takes non-zero values for density
contributions located inside a circle of radius $\sigma$ around its maximum. 
Fig.~\ref{3dcompactn} quantifies the above observations. In the strong
Boltzmann spinning regime ($C = 10$, left) the blob retains much of its
initial shape, so that the compactness is higher compared to the weak
Boltzmann spinning regime ($C = 100$, right), where filaments feature a more 
bean-shaped structure which reduces the compactness measure. 
At the time of measurement ($t = 5$), the heavier isotopic blobs show slightly more
compactness, which is an indirect result of decreased velocity: at a given
time, the lighter isotopic blobs are further developed and thus less circular. 
The observed trends are similar for cold ($\tau_i=0$) and warm ($\tau_i=1$) ions.

For increased collisionality, the deviation from circularity is more pronounced, as
the mushroom-cape shape is realized. Blobs in light isotopic plasmas are then
again further  developed, i.e. finer scales have emerged at the time of
recording, resulting in a sharper mass dependence of blob compactness compared
to $C = 10$, where smaller scales are less prominent. 

\begin{figure} 
  \begin{subfigure}{8cm}
    \centering\includegraphics[width=7cm]{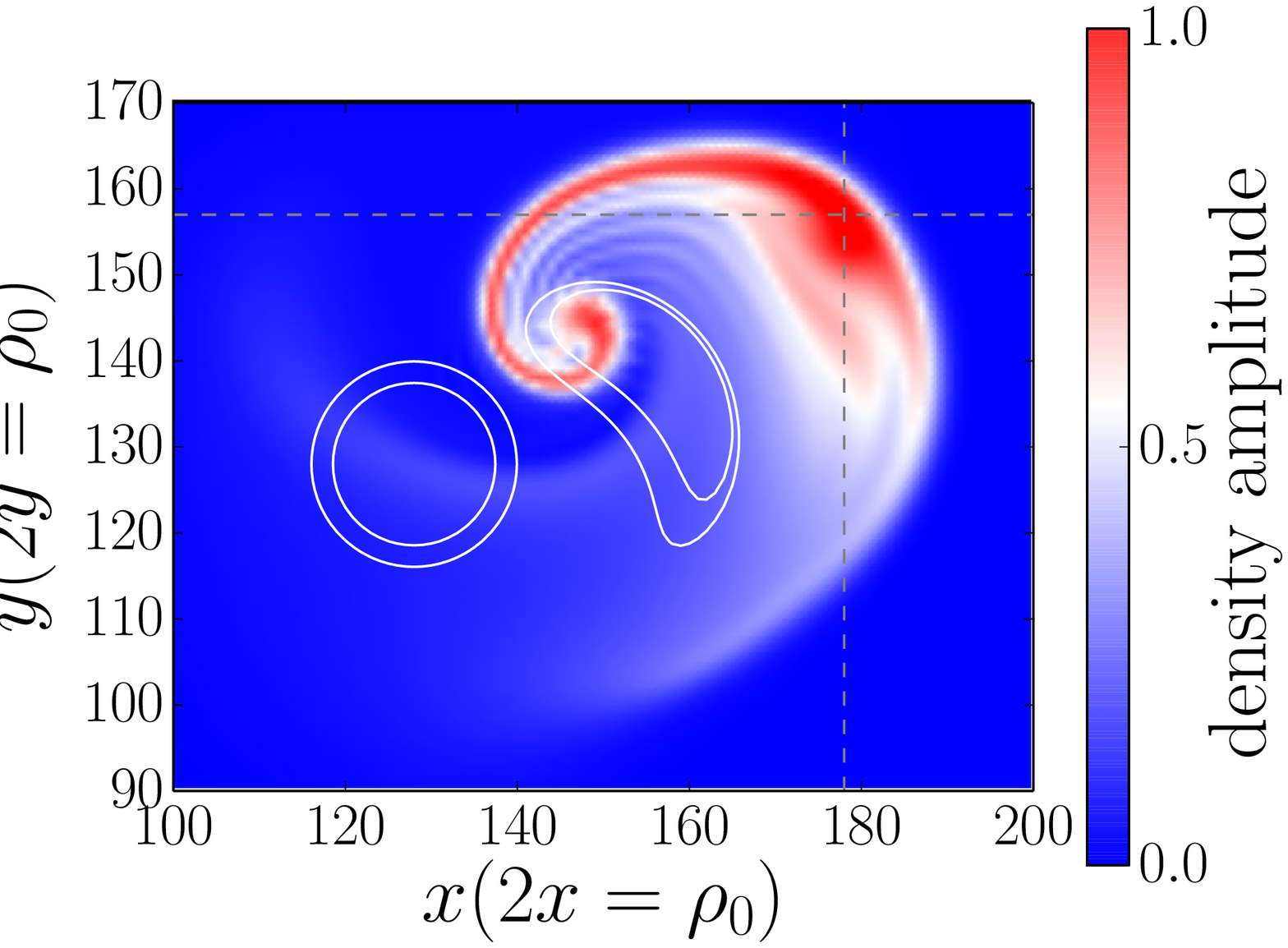} 
  \end{subfigure}
  \begin{subfigure}{8cm}
    \centering\includegraphics[width=7cm]{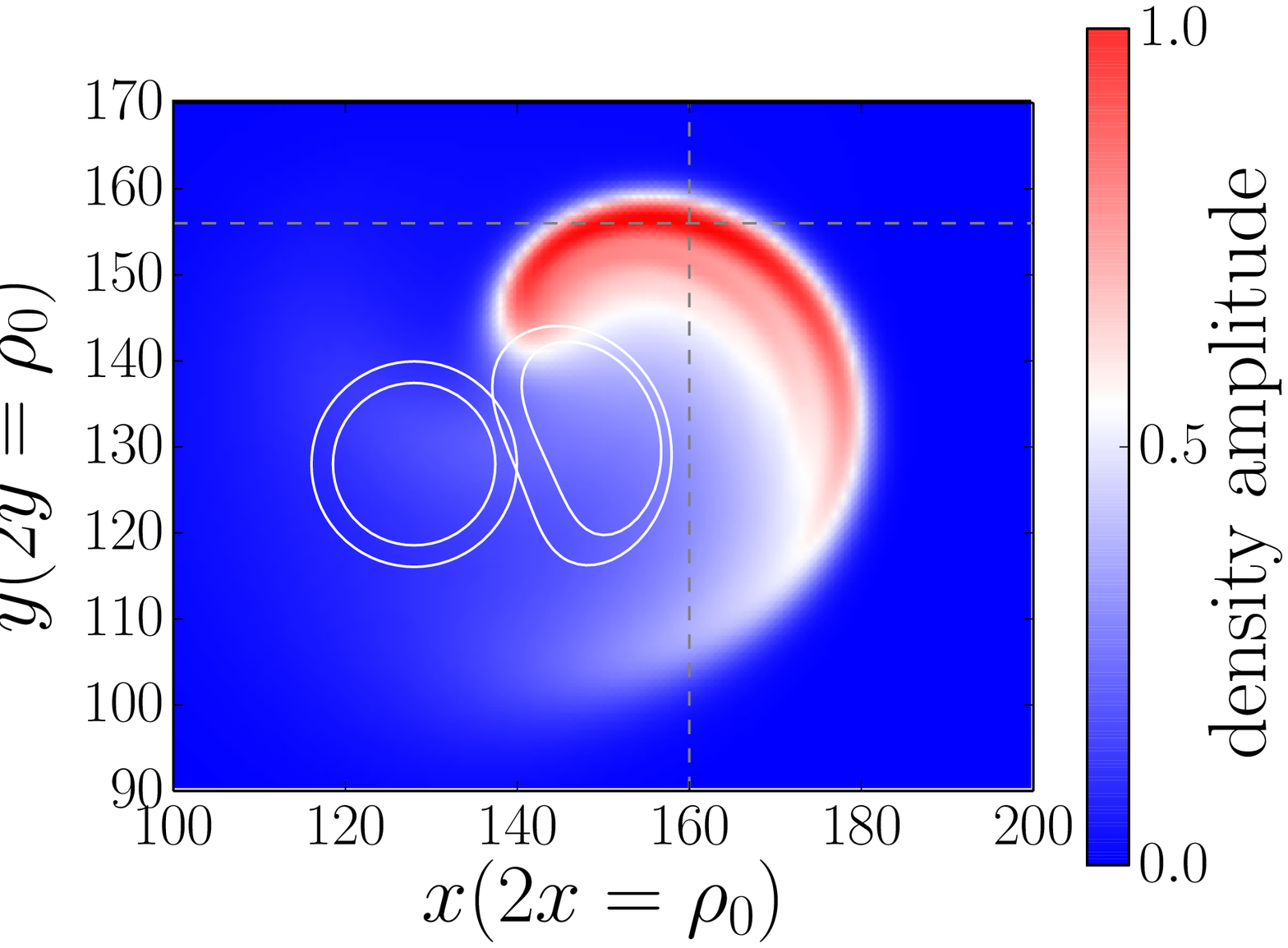} 
  \end{subfigure}
  
  \begin{subfigure}{8cm}
    \centering\includegraphics[width=7cm]{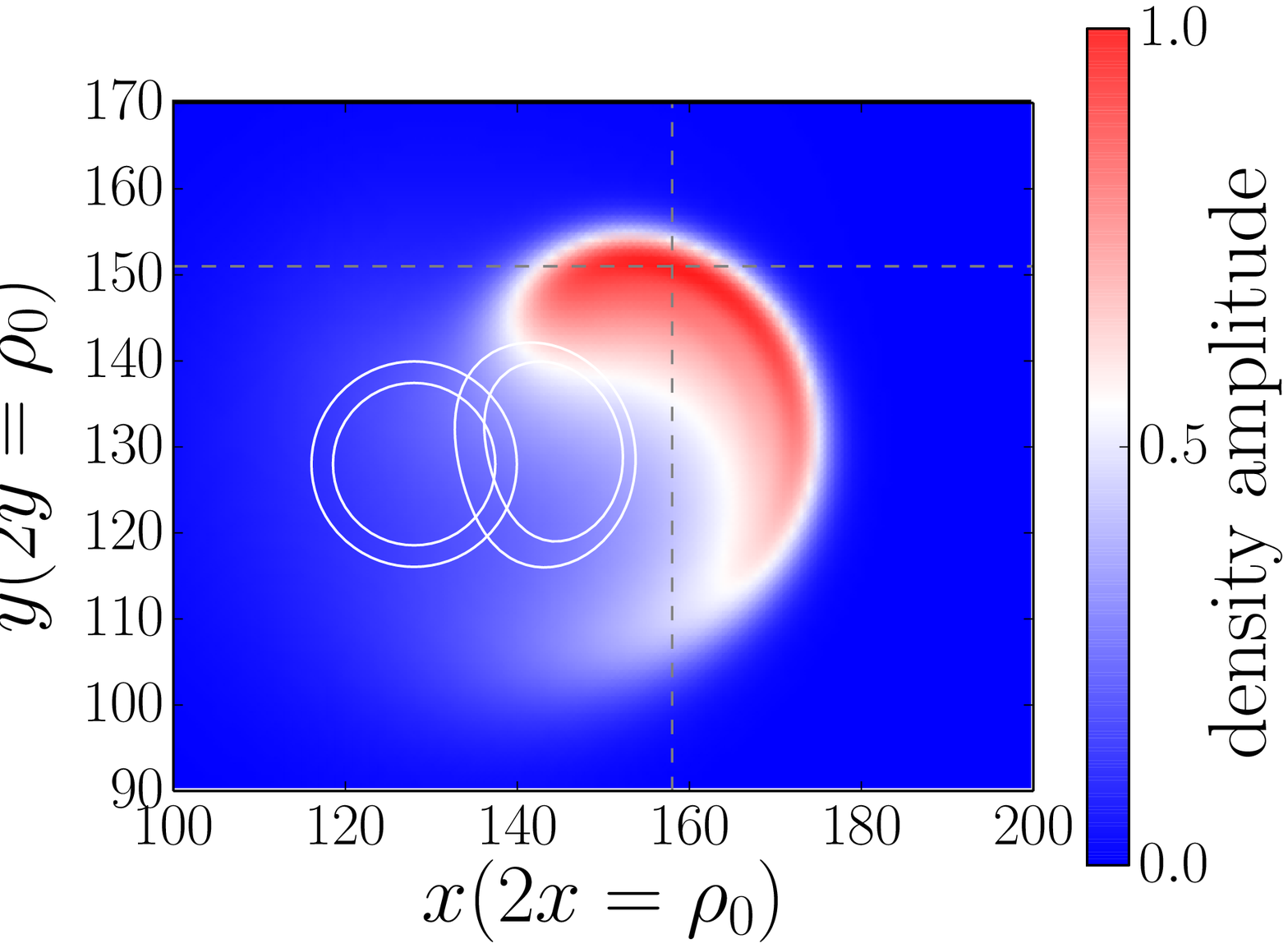} 
  \end{subfigure}
  \begin{subfigure}{8cm}
    \centering\includegraphics[width=7cm]{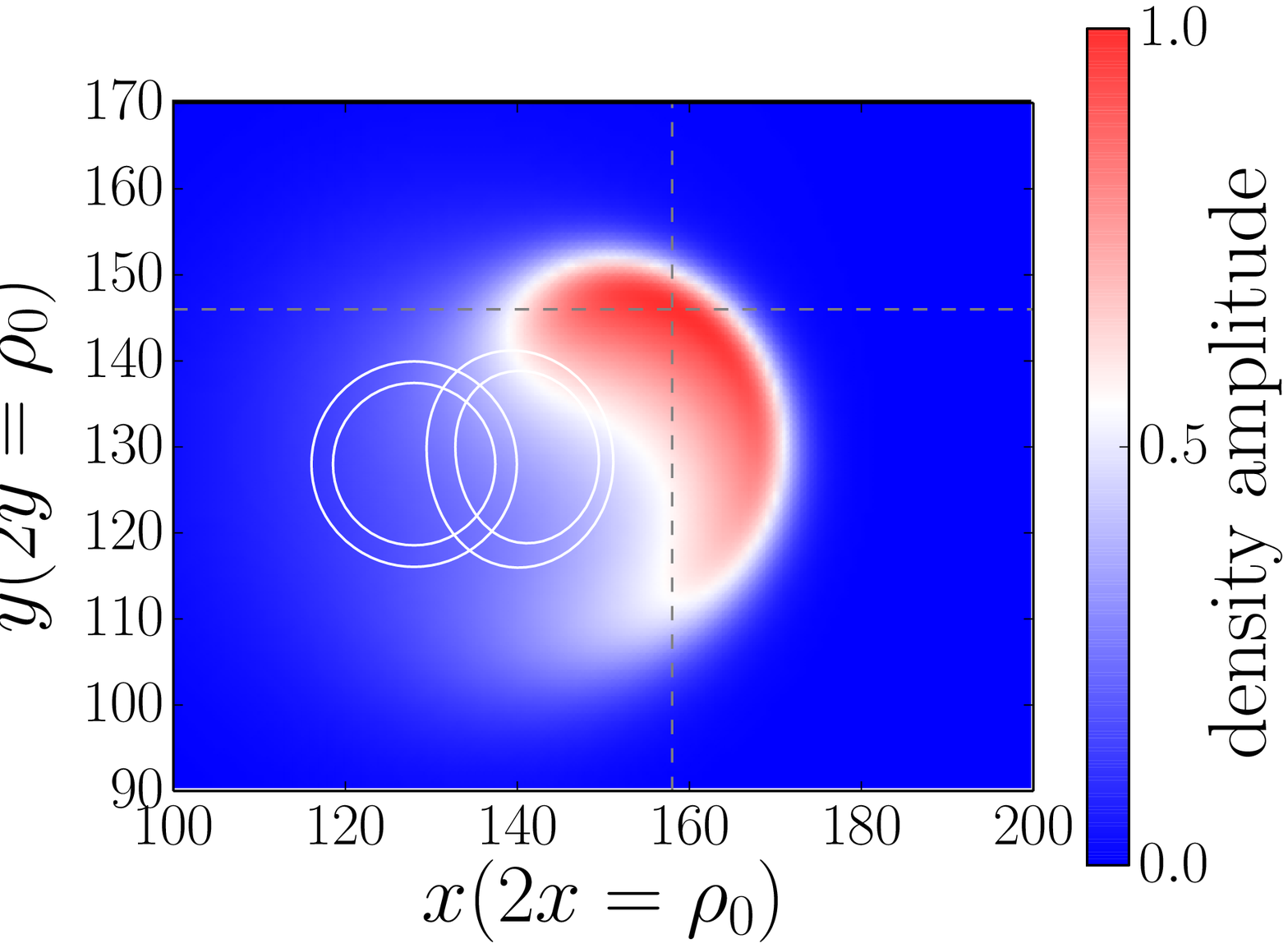} 
  \end{subfigure}
\caption{Density contour plots of cold ion blob propagation for different
  plasma species and $C = 100$. White contours depict the initial state ($t =
  0$) and a subsequent snapshot at $t = 4$.  The color plot shows the outboard
  midplane ($z=z_0$) electron density perturbation at $t = 8$. Shown are
  protium (top left), deuterium (top right), tritium (bottom 
left) and singly charged helium-4 (bottom right). The dotted lines give the
$x-y$ coordinates of the density maximum.}   
\label{3dcontt0}
\end{figure}

\begin{figure} 
  \begin{subfigure}{8cm}
    \centering\includegraphics[width=7cm]{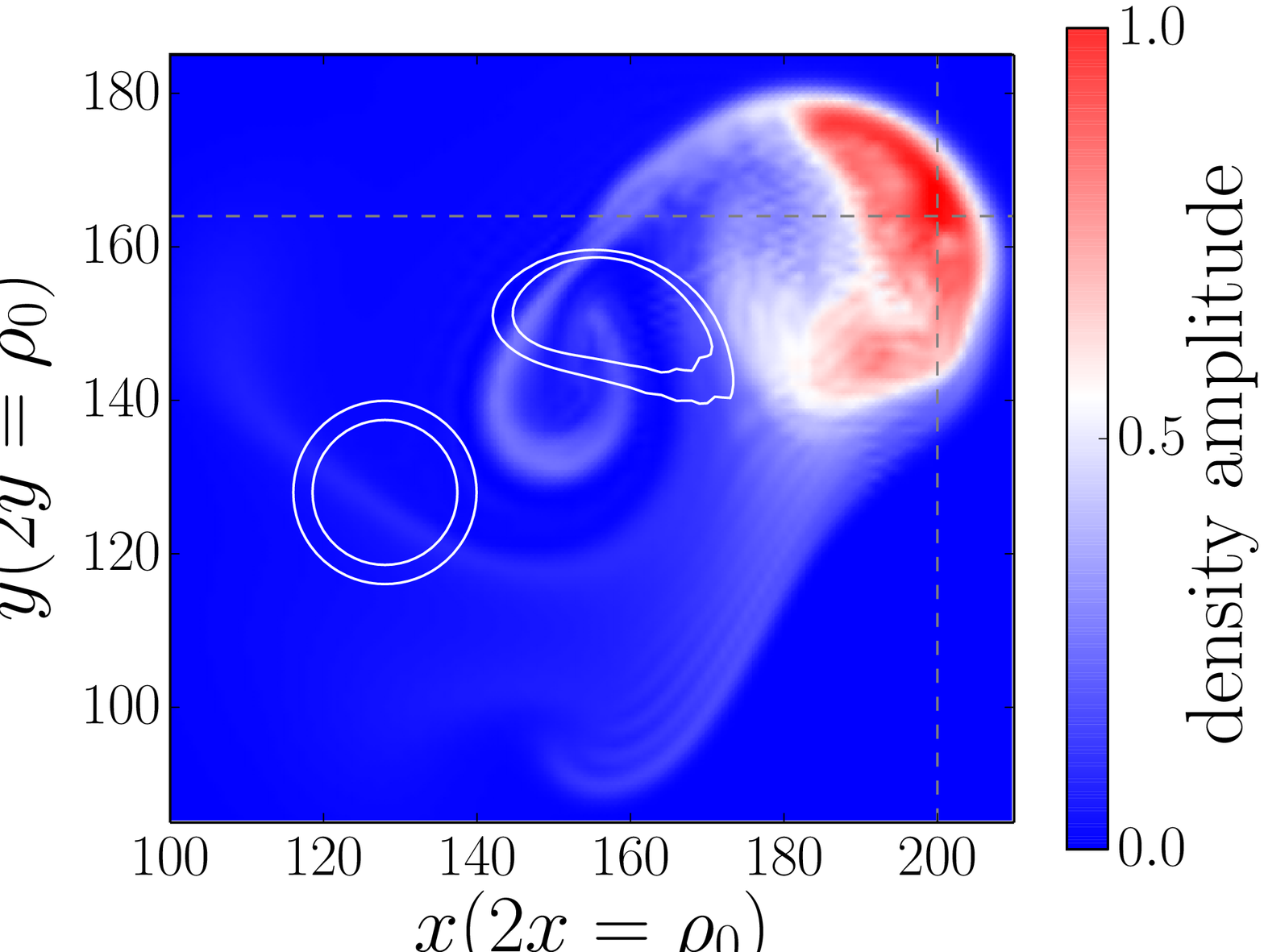} 
  \end{subfigure}
  \begin{subfigure}{8cm}
    \centering\includegraphics[width=7cm]{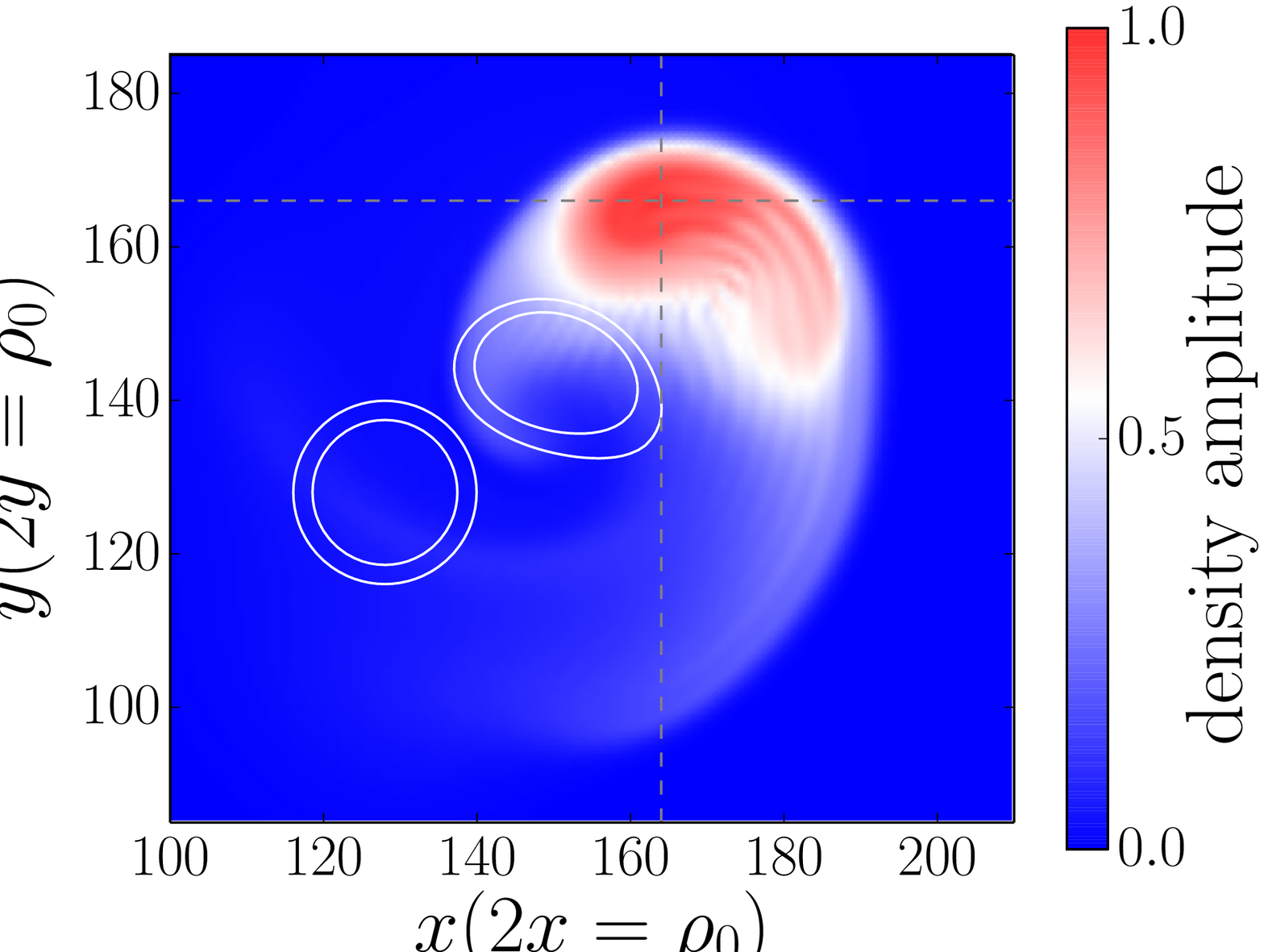} 
  \end{subfigure}
  
  \begin{subfigure}{8cm}
    \centering\includegraphics[width=7cm]{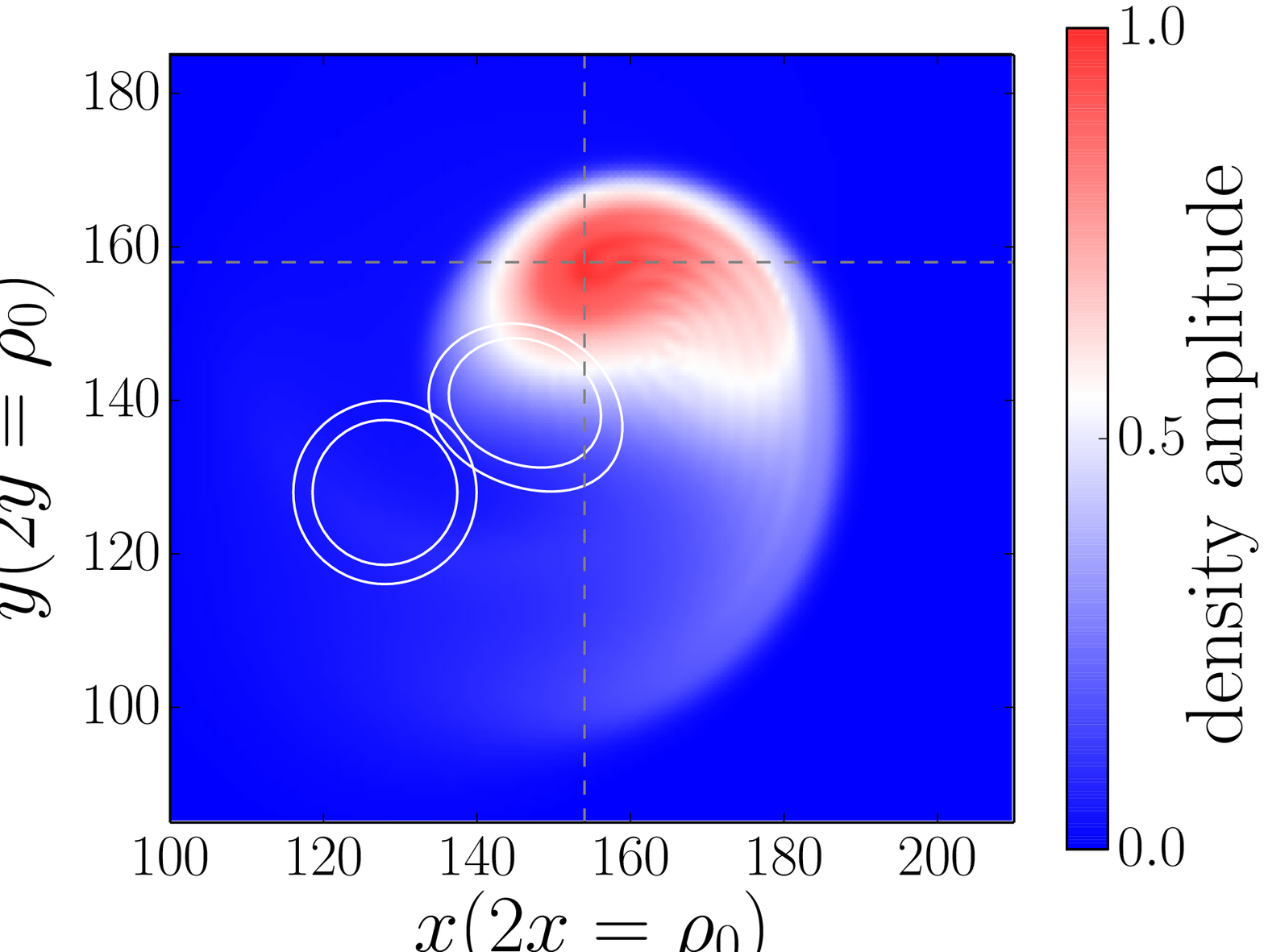} 
  \end{subfigure}
  \begin{subfigure}{8cm}
    \centering\includegraphics[width=7cm]{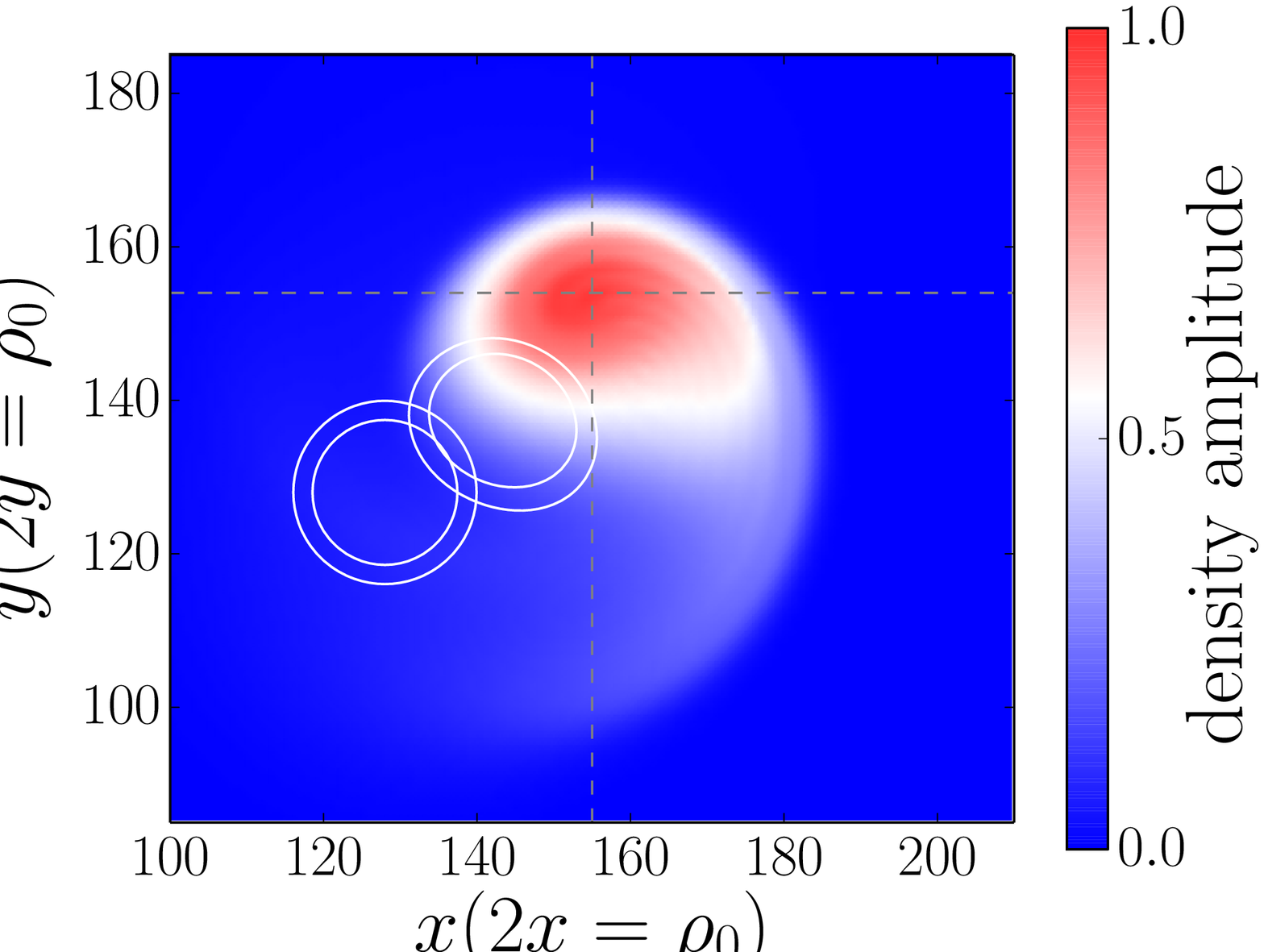} 
  \end{subfigure}
\caption{Density contour plots of warm ion blob propagation for different
  plasma species and $C = 100$. White contours depict the initial state ($t =
  0$) and a subsequent snapshot at $t = 4$.  The colour plot shows the
  outboard midplane ($z=z_0$) electron density field at $t = 8$. Shown are
  protium (top left), deuterium (top right), tritium (bottom 
left) and singly charged helium-4 (bottom right). The dotted lines give the
$x-y$ coordinates of the density maximum.}   
\label{3dcontt1}
\end{figure}

\begin{figure} 
  \begin{subfigure}{8cm}
    \centering{\includegraphics[width=7cm]{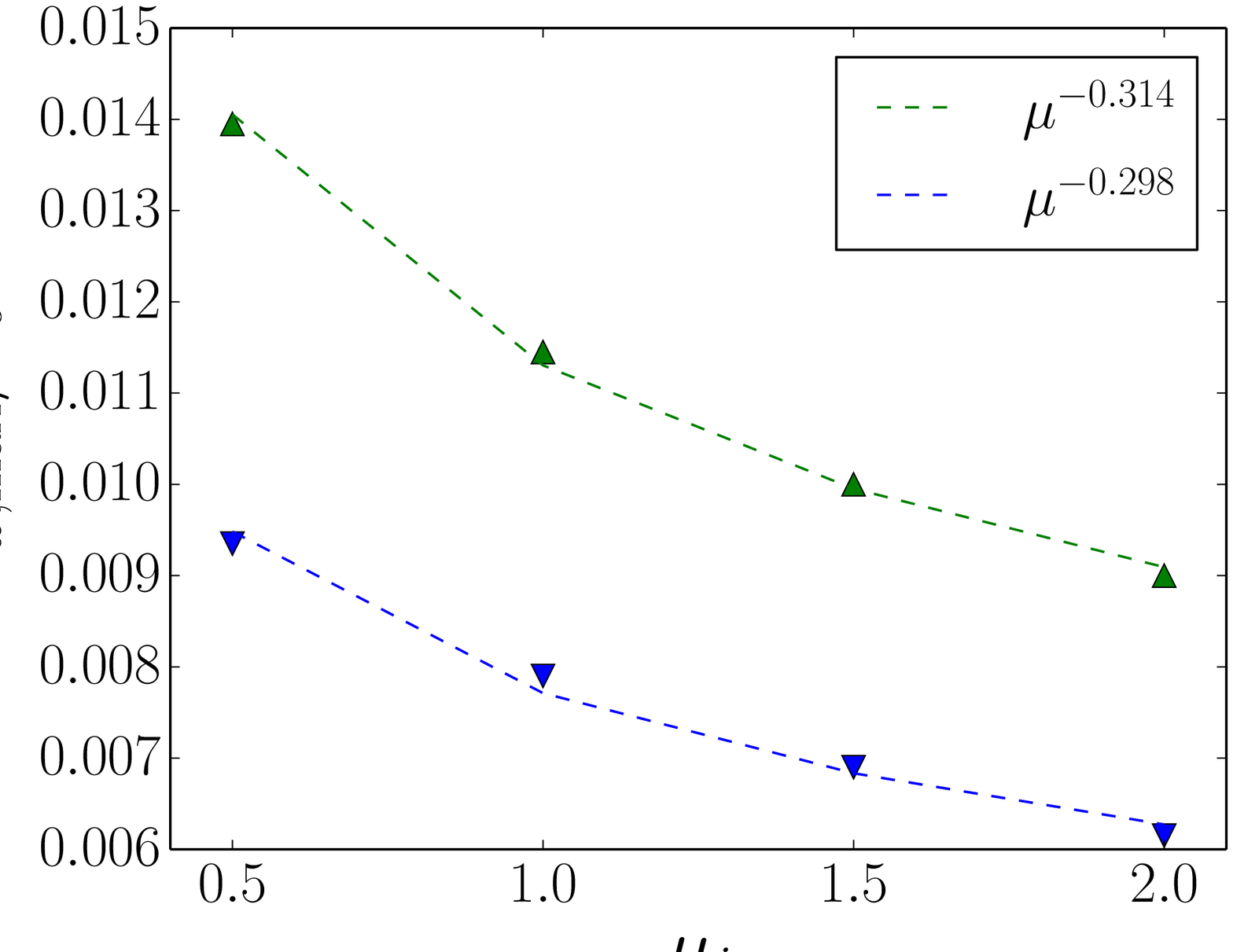}} 
  \end{subfigure}
  \begin{subfigure}{8cm}  
    \centering{\includegraphics[width=7cm]{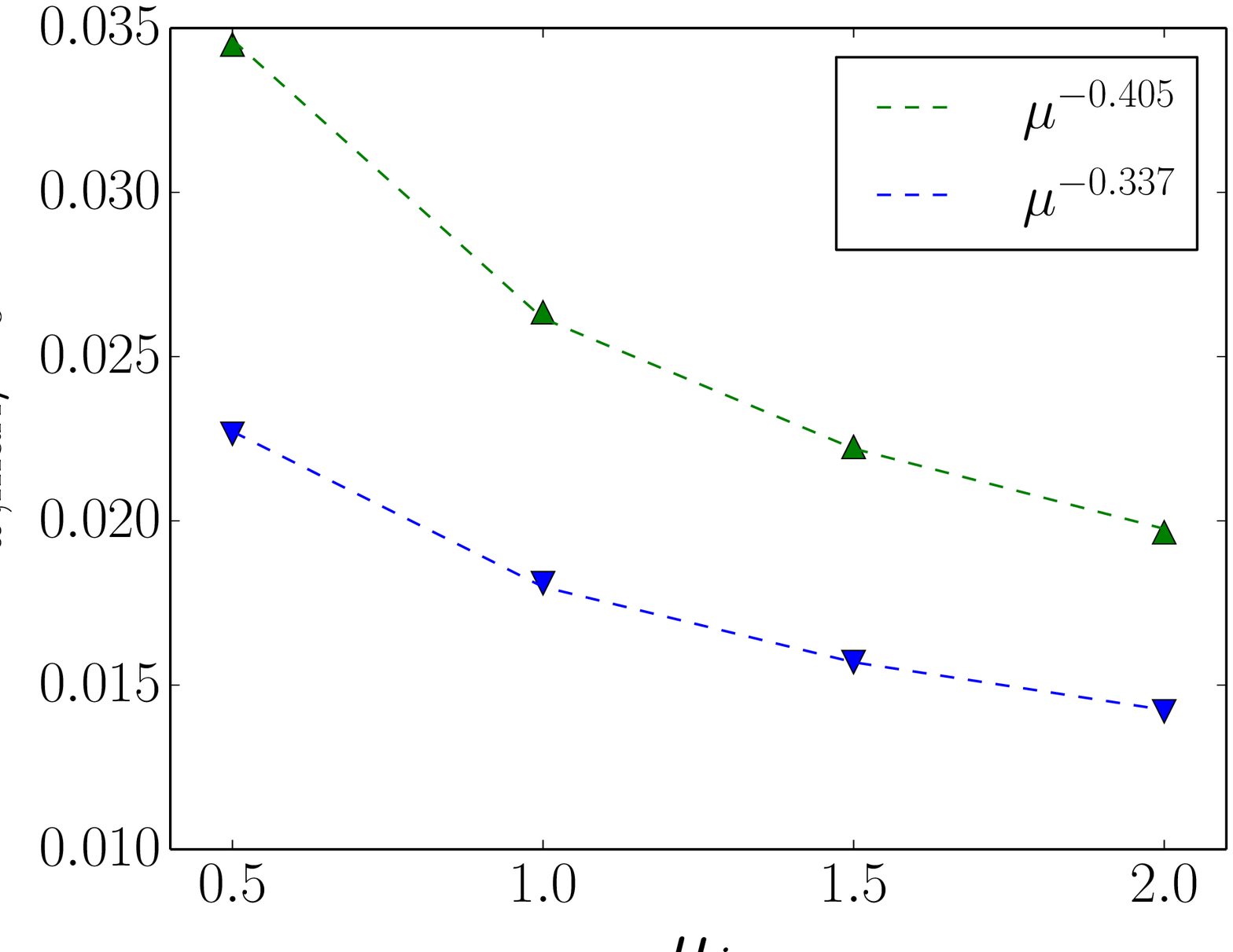}}
  \end{subfigure}
\caption{Maximum radial center-of-mass velocities for strong Boltzmann spinning
  ($C = 10$, left) and reduced spinning ($C = 100$, right) for cold (blue) and
  warm ions (green), using zero vorticity initialisation.}  
\label{zgf}
\end{figure}

Fig.~\ref{3dcontt0} shows filament propagation for cold ions ($\tau_i=0$) and weak
Boltzmann spinning. This can be compared to Fig.~\ref{3dcontt1} which shows
propagation for warm ions ($\tau_i=1$) and also weak Boltzmann spinning. 
It is observed that there is poloidal propagation also for the cold ion case,
which is a consequence of the non-vanishing Boltzmann spinning that is also present, although greatly reduced, for these high collisionality ($C = 100$) cases. 
The resulting maximum radial center-of-mass velocites at the outboard midplane
($z = z_0$) are shown in Fig.~\ref{zgf} for weak (right) and strong (left)
Boltzmann spinning.

\begin{figure} 
    \centering{\includegraphics[width=8cm]{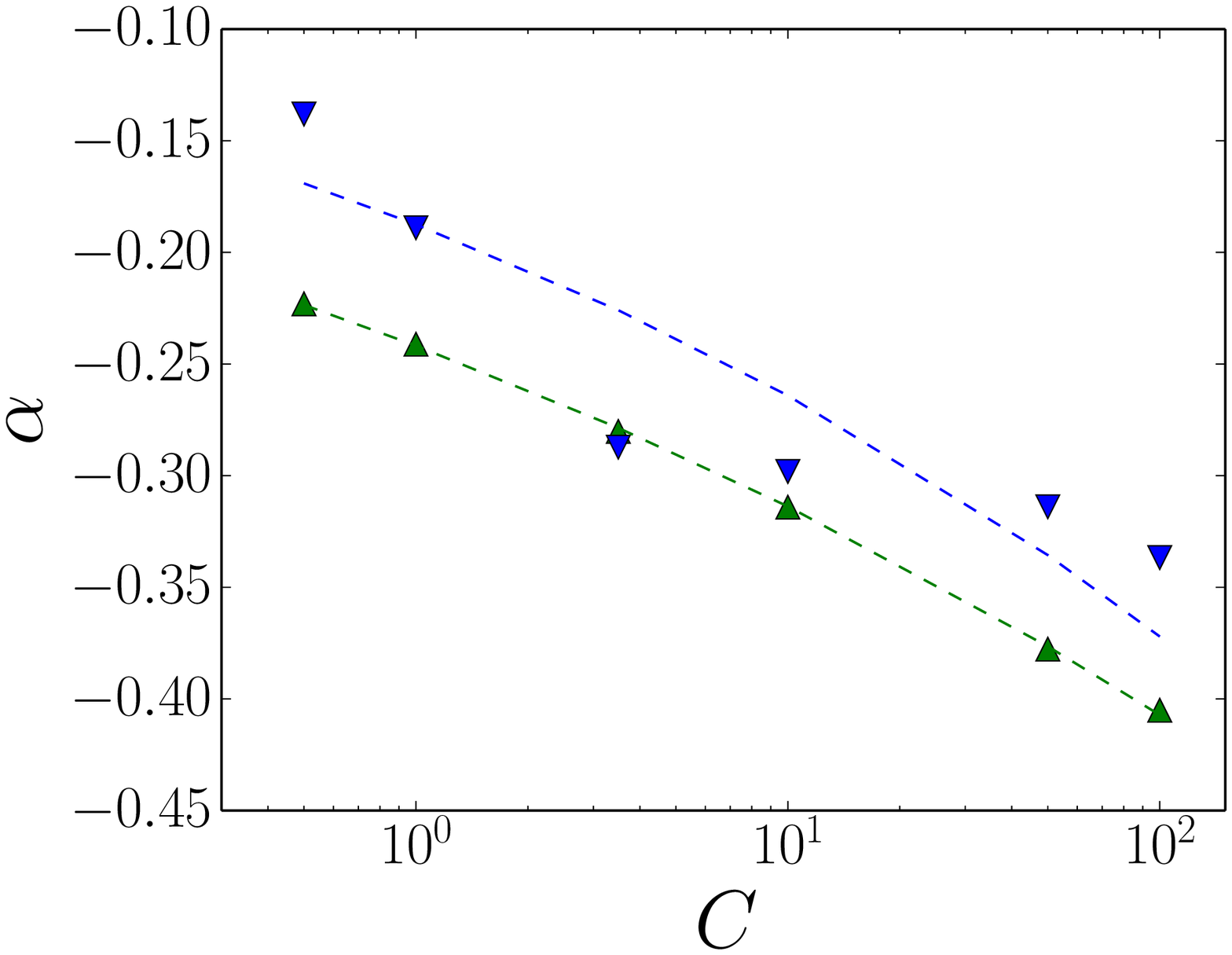}} 
\caption{Maximum radial velocity scaling exponents $\alpha$ (analysed at $z = z_0$) as
  a function of the collisionality parameter $C$ for cold (blue) and warm ions
  (green). The scaling exponents with respect to collisionality are $\alpha
  \sim C^{0.15}$ (cold ions) and $\alpha \sim C^{0.11}$ (warm ions).}  
\label{sc_exp}
\end{figure}

The fits of the exponent in $\mu^\alpha$ to the simulation data in
Fig.~\ref{zgf} carry evidence that the additional mass dependences
introduced by the 3-d model via parallel sheath-boundary conditions and
parallel ion velocity dynamics causes the clear deviation from a $1 /
\sqrt{\mu_i}$ scaling.  

For high collisionality (and thus reduced Boltzmann spinning), the parallel
current is impeded and the dynamics is more two-dimensional than for lower
collisionalities. The competing nature of the parallel divergence versus
current continuity via the divergence of the polarisation current with
collisionality  is shown in Fig.~\ref{sc_exp}:
for each value of collisionality $C$ we compute the isotopic dependence of the
outboard-midplane maximum center-of-mass radial velocity, 
\begin{eqnarray}
V_\mathrm{max} \sim \mu_i^{\alpha (\mu_i)},
\end{eqnarray}
contained in the scaling exponent, $\alpha (\mu_i)$. 
For large values of $C$ the resulting dynamics strongly features
2-d propagation characteristics, since the diamagnetic current is almost
exclusively closed via the polarisation current, which gives the 
$1 / \sqrt{\mu_i}$ scaling introduced in sec.~\ref{two_d}. 
Note that the scaling with respect to $C$ cannot be inferred from linear models.

\subsubsection{Non-zero gyrofluid vorticity initialisation} 
\label{nozGF}

So far we have in this section applied the initial condition $n_\rme \equiv \Gamma_1 n_i$
associated with zero initial vorticity. Now the case for non-zero initial
vorticity by the condition $n_\rme = n_i$ is considered. 

\begin{figure} 
  \begin{subfigure}{8cm}
    \centering{\includegraphics[width=7cm]{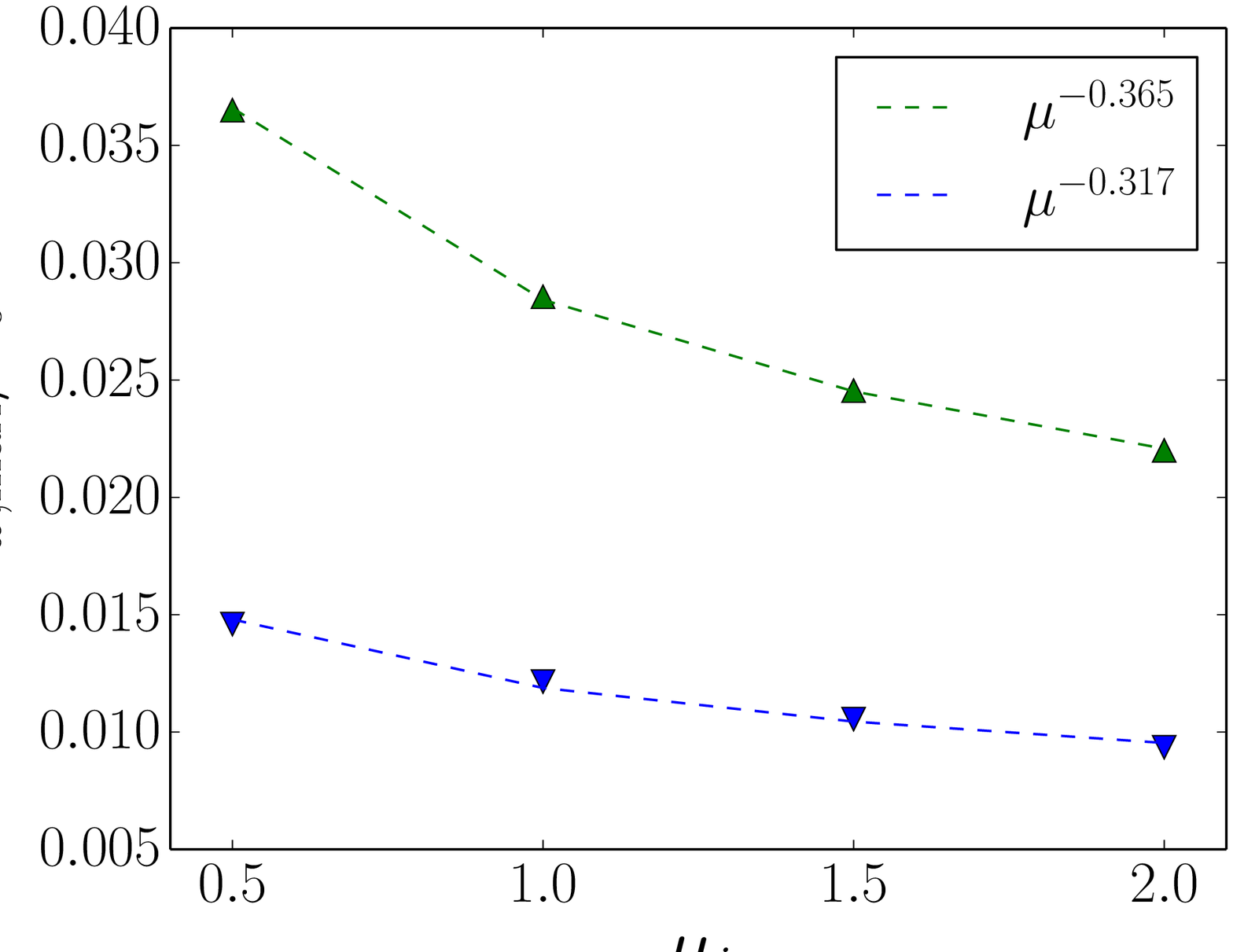}} 
  \end{subfigure}
  \begin{subfigure}{8cm}  
    \centering{\includegraphics[width=7cm]{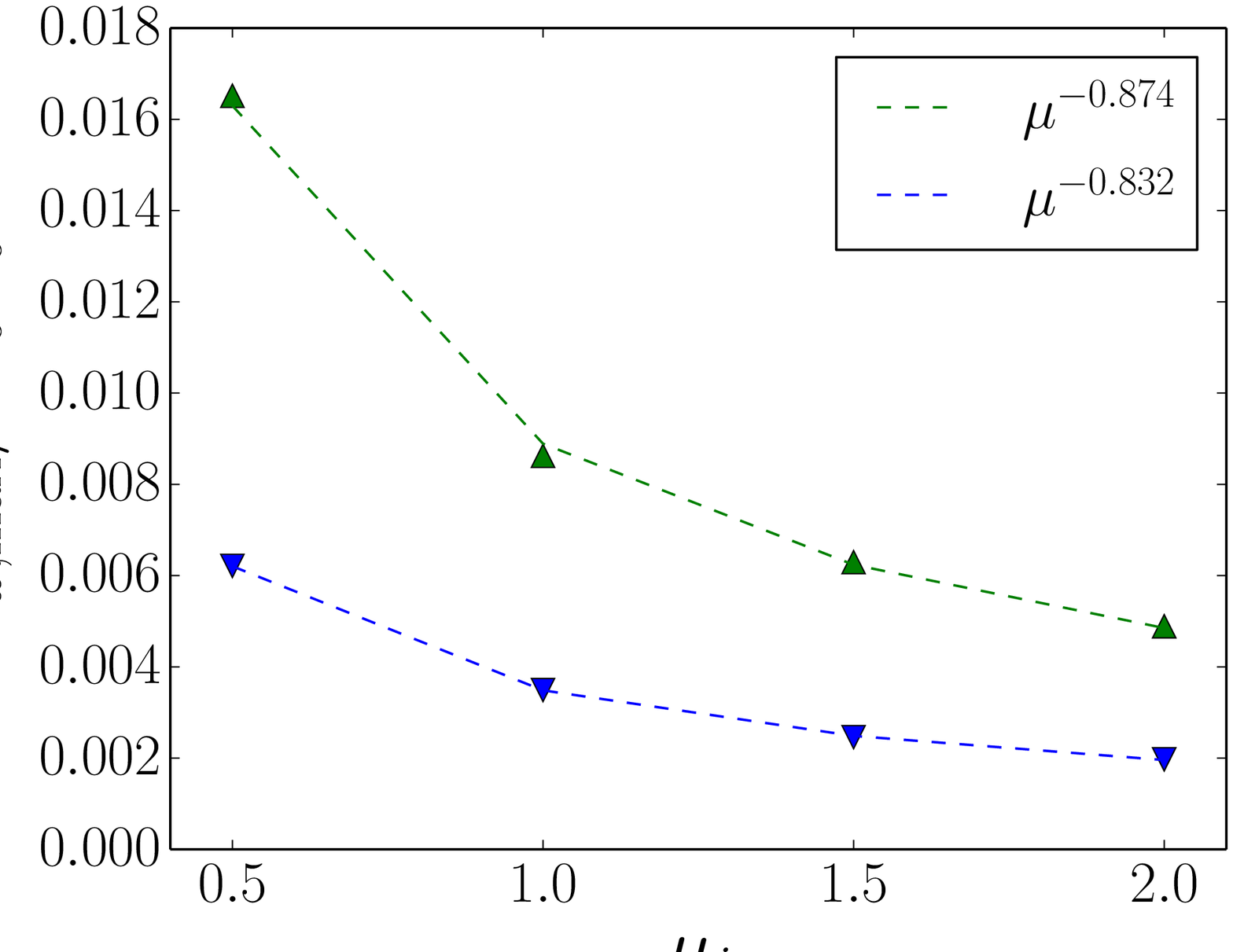}}
  \end{subfigure}
\caption{Maximum radial center-of-mass velocity (left) and radial acceleration
  (right) at $z=z_0$ for 3-d warm ion blobs, using non-zero vorticity initialisation at each
  drift plane, for $C = 10$ (blue) and $C = 100$ (green).}    
\label{nzgf}
\end{figure}

Fig.~\ref{nzgf} shows results for warm ion ($\tau_i = 1$) computations in the strong  
($C = 10$, blue) and weak ($C = 100$, green) Boltzmann spinning regimes.  
(Recall that for $\tau_i = 0$ this discussion is redundant since $\Gamma_1 (\tau_i = 0)= 1$.)
The left figure depicts the maximum radial center-of-mass velocity, and the right
figure shows the corresponding average acceleration.

Comparing with Fig.~\ref{zgf} we notice that the resulting filament velocities
are similar to those obtained from zero initial vorticity. We also
find that the isotopic dependence $\sim \mu_i^\alpha$ is not significantly altered.

Recalling the results from 2-d computation in sec.~\ref{two_d}, we may
conclude that the initialisation is not that important for 3-d numerical
simulations with respect to the maximum radial filament velocity. 
The slight impact of the initial condition on the resulting scaling exponent
for the 2-d case may then be connected to the more prominent mass
dependence in the polarisation current, which is weakened when
parallel currents are taken into account.

\subsubsection{Comparison of filaments in deuterium and helium plasmas} 
\label{DeutvHe}

When comparing blob filament propagation in deuterium and in fully ionised
helium-4 plasmas in the present model, the dynamical evolution is identical in
the cold ion limit: in the model parameter $\mu_{i} = m_{i} /(Z_i m_{D})$ the
doubled mass of the helium nucleus excactly cancels with the doubled positive
charge, $Z_\mathrm{He} = 2$. 

Differences are only appearing in warm ion cases. The normalised 
mass ratio is now identical for both species, $\mu_\mathrm{D} = \mu_\mathrm{He} =
1$. The only model parameter that is different, is the helium temperature
ratio, $\tau_\mathrm{He} = T_\mathrm{He} / Z_\mathrm{He} T_\rme$. The species mass
effects thus appears in the combined $b \sim \mu_i \tau_i$.
In the following we consider plasmas at equal temperature, $T_\mathrm{D} =
T_\mathrm{He} = 2 T_\rme$ such that $\tau_\mathrm{D} = 2$ and $\tau_\mathrm{He} = 1$. 

The higher charge state of the helium nucleus is also indirectly evident in
the reduced electron-ion collision frequency contained in the $C$ -parameter. 
For electron-ion collisions where the ions are in charge state $Z_i$ we have $C
\sim \alpha_e \nu_\rme$, with \cite{Hirshman} 
\begin{equation} \label{c_dep}
\alpha_e \approx \frac{1 + 1.198 Z_i + 0.222 Z_i^2}{1 + 2.966 Z_i + 0.753 Z_i^2}.
\end{equation}
For $Z_i = 1$ we have $\alpha_e \approx 0.51$ and $Z_i = 2$ gives $\alpha_e
\approx 0.43$. 

To account for this dependence, we in the following consider two cases: 
(1) equal non-normalised collision frequencies, i.e. $C \rightarrow 0.51 C$
for deuterium and $C \rightarrow 0.43 C$ for helium; (2) using the same $C$
for both deuterium and helium computations.  

In case (2), setting first the normalised collisionality parameter $C = 10$
identical for both D and He computations results in $V_\mathrm{D} = 1.41
\delta c_{s0}$ and $V_\mathrm{He} = 1.14 \delta c_{s0}$. 
Setting $C = 100$ identical for both D and He gives $V_\mathrm{D} = 3.2 \delta
c_{s0}$ and $V_\mathrm{He} = 2.6 \delta c_{s0}$, respectively. 

For case (1) we set the electron-ion collision frequency equal for both
species, so that different $C$ parameters are used according to equation \ref{c_dep}:
$C_\mathrm{D} = 10$ corresponds to $C_\mathrm{He} = 8.43$, and 
$C_\mathrm{D} = 100$ to $C_\mathrm{He} = 84.3$. 
In these cases, maximal radial He velocities are $V_\mathrm{He} (C=10) = 1.09
\delta c_{s0}$ and $V_\mathrm{He} (C=100) = 2.46 \delta c_{s0}$. 

\begin{figure} 
  \begin{subfigure}{8cm}
    \centering{\includegraphics[width=7cm]{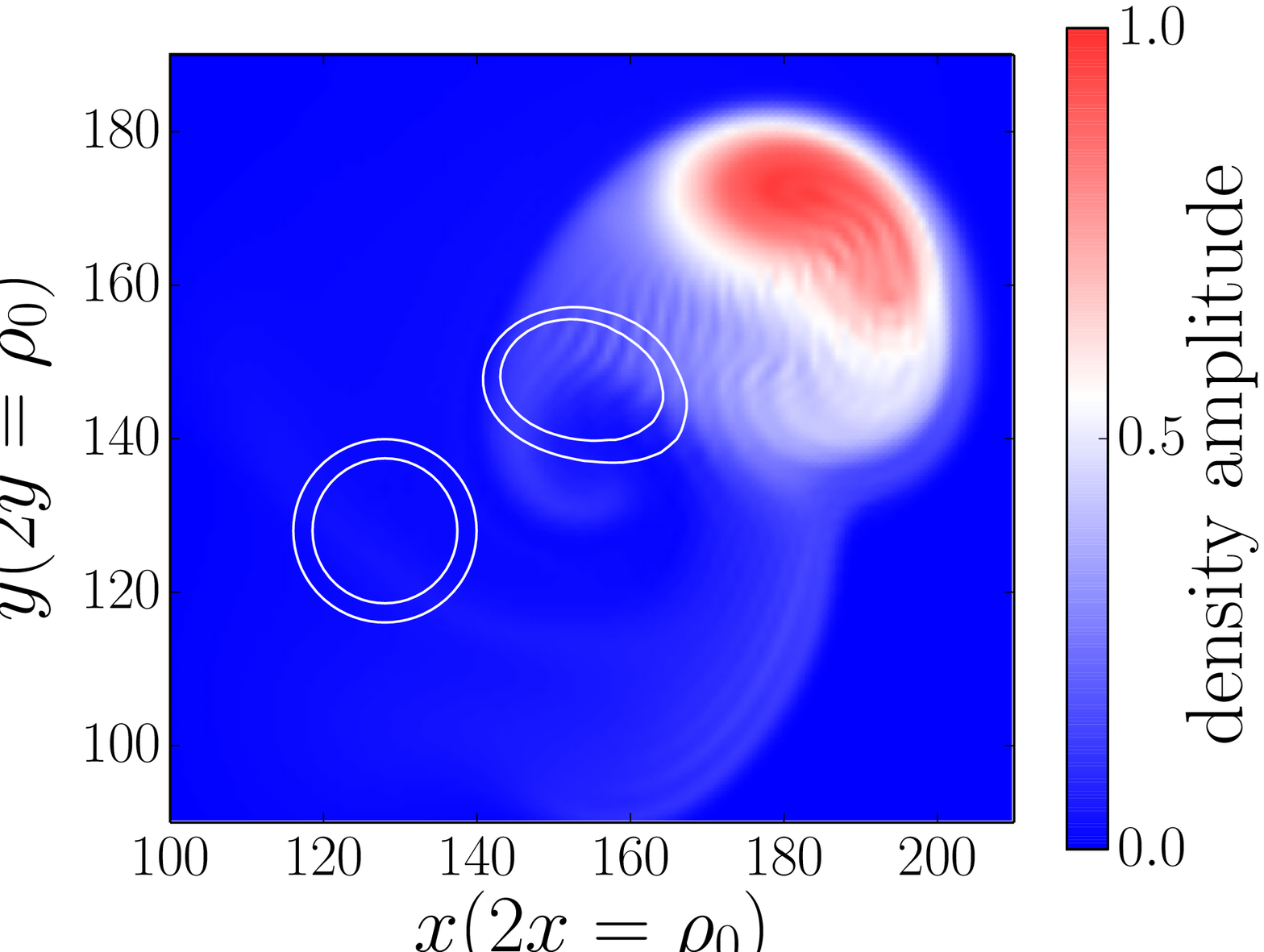}} 
  \end{subfigure}
  \begin{subfigure}{8cm}  
    \centering{\includegraphics[width=7cm]{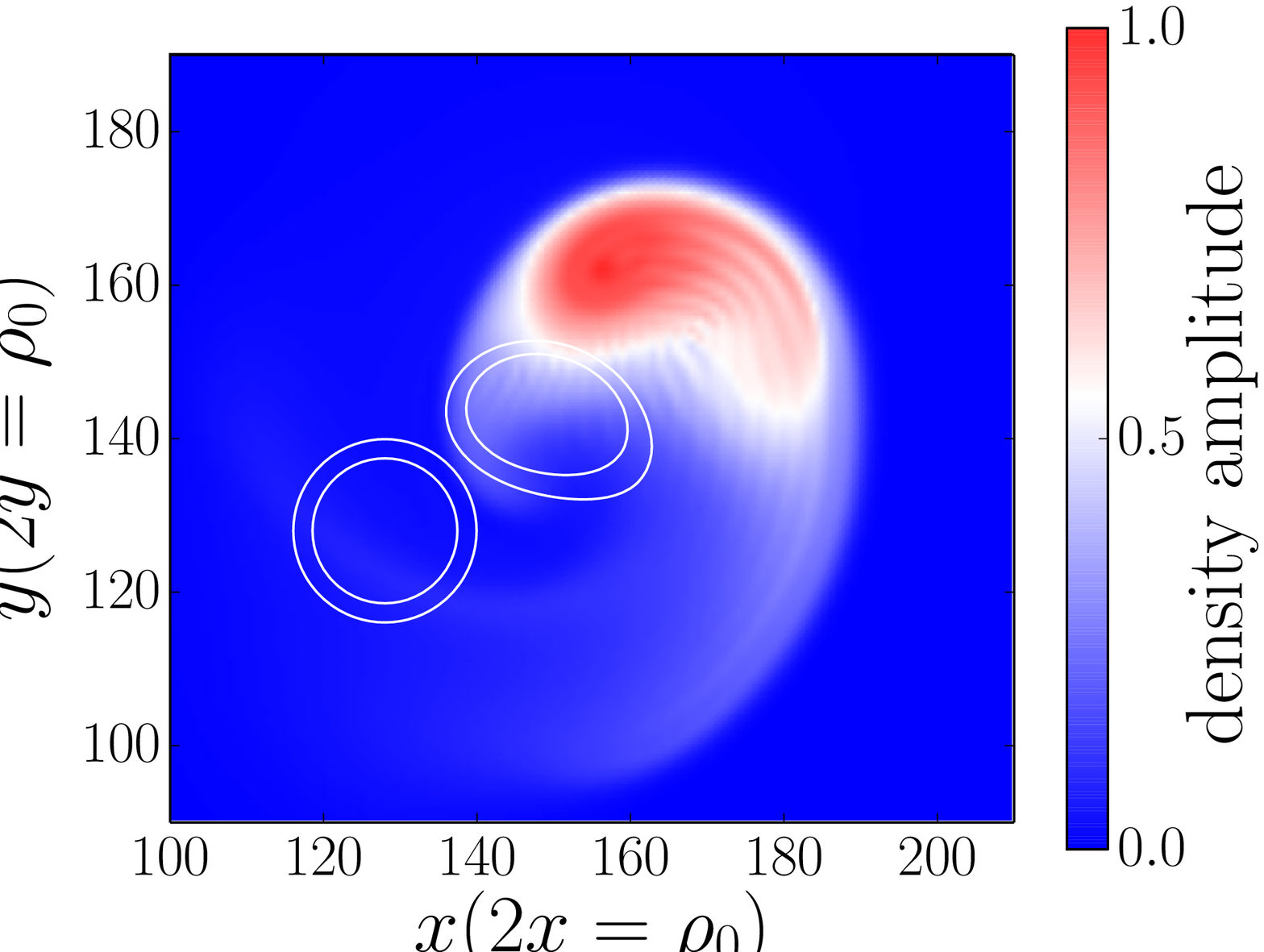}}
  \end{subfigure}
\caption{Density contour plots of filament cross sections in deuterium (left)
  and fully ionised helium-4. White contours depict the initial state ($t =
  0$) and a subsequent snapshot at $t = 4$. The colour plot shows the
  outboard-midplane ($z=z_0$) electron density perturbation field at $t =
  8$. In both cases $T_i = 2 T_\rme$  so that $\tau_\mathrm{D} = 2$ and
  $\tau_\mathrm{He} = 1$. Electron-ion collision frequencies are set equal
  resulting in $C_\mathrm{D} = 100$ and $C_\mathrm{He} = 84.3$.}  
\label{DvHe}
\end{figure}

We find that regardless of how the charge state depency for the  
relative value of the collision parameter is treated, the filaments in
deuterium plasmas move faster than in helium plasmas at identical temperature. 
This is visualized in Fig.~\ref{DvHe} showing filament propagation at equal
electron-ion collision frequency and electron temperature. 

\section{Conclusions} 
\label{conclusion}

We have investigated filament propagation in SOL conditions characteristic for
tokamak fusion devices. Quasi-2-d dynamics is restored in high resistivity
regimes, where the maximum radial blob velocity scales inversely proportional
with the square root of the ion mass. 
In 2-d the diamagnetic current drive is closed solely via the polarisation current,  
yielding this simple characteristic scaling. 

The larger inertia through polarisation of more massive ion species
effectively slows the evolution of filaments, and the maximum radial velocity
occurs later compared to blobs in plasmas with lighter ions.

For non-zero initial vorticity condition, the 2-d warm ion blobs show compact
radial propagation, where the isotopic effect through the mass dependent
FLR terms is slightly less pronounced.

Boltzmann spinning appears in 3-d situations particularly for low
collisionality regimes, and leads to a reduced dependence on the ion isotope
mass. The exponent in the scaling $V \sim \mu_i^\alpha$ has been found to be
typically within the range $\alpha \in [-0.1, -0.3] $ for $C < 10$, which is a
regime relevant for the edge of most present tokamaks. 

Considering current continuity, the closure via the parallel current
divergence dynamically competes with current loops being closed through the
polarisation current.
For high collisionalities the parallel current is effectively imepeded and 
the polarisation current characteristics dominate the blob evolution,
producing a more 2-d like velocity dependence with respect to ion mass. 
The initial condition has been found to have little influence on the maximum
radial velocity when in 3-d the parallel closure of the current is taken into account. 

For similar ion temperatures and electron-ion collision frequencies, it has
been found that helium filaments travel more slowly compared to deuterium
filaments in both high and low collisionality regimes.

This work was devoted to the identification of isotopic mass effects on
seeded (low amplitude) blob filaments in the tokamak SOL by means of a
delta-$f$ gyrofluid model. 
Naturally, blobs emerge near the separatrix within coupled edge/SOL
turbulence. The dependence of fully turbulent SOL transport on the ion mass
therefore will have to be further studied within a framework that consistently
couples edge and SOL turbulence, preferably through a full-$f$ 3-d gyrofluid
(or gyrokinetic) computational model that does not make any smallness
assumption on the relative amplitude or perturbations compared to the
background \cite{Matthias,Held16,Kendl15}.

\section*{Acknowledgments}
We acknowledge main support by the Austrian Science Fund (FWF) project Y398.
This work has been carried out within the framework of the EUROfusion
Consortium and has received funding from the Euratom research and training
programme 2014-2018 under grant agreement No 633053. The views and opinions
expressed herein do not necessarily reflect those of the European Commission.  


\end{document}